\documentclass[preprint]{aastex62}

\usepackage{array}
\usepackage{multirow}
\usepackage[utf8]{inputenc}
\graphicspath{{./}{figures/}}
\newcommand{\beq}{\begin{equation}}
\newcommand{\eeq}{\end{equation}}

\shorttitle{Identifying Binary \textit{Kepler} Exoplanet Host Stars}
\shortauthors{Colton et al.}
\begin{document}
\title{Identifying Bound Stellar Companions to \textit{Kepler} Exoplanet Host Stars Using Speckle Imaging}

\correspondingauthor{Elliott Horch}
\email{horche2@southernct.edu}

\author{Nicole M. Colton}
\altaffiliation{Current Address: Department of Physics, 1875 Campus Delivery, Colorado State University, Fort Collins, CO 80523}
\author{Elliott P. Horch}
\altaffiliation{Adjunct Astronomer, Lowell Observatory; Visiting Astronomer, Kitt Peak National Observatory;
Visiting Astronomer, Gemini Observatory}
\affiliation{Department of Physics, Southern Connecticut State University, 501 Crescent Street, New Haven, CT 06515, USA}

\author{Mark E. Everett}
\altaffiliation{Visiting Astronomer, Gemini Observatory}
\affiliation{NSF's National Optical-Infrared Astronomy Research Laboratory, 950 North Cherry Avenue, Tucson, AZ 85719, USA}

\author{Steve B. Howell}
\altaffiliation{Visiting Astronomer, Kitt Peak National Observatory; Visiting Astronomer, Gemini Observatory}
\affiliation{NASA Ames Research Center, Moffett Field, CA 94035 USA}

\author{James W. Davidson, Jr.}
\altaffiliation{Visiting Astronomer, Lowell Observatory}
\affiliation{Department of Astronomy, University of Virginia, P.O. Box 400325, Charlottesville, VA 22904, USA}

\author{Brian J. Baptista}
\affiliation{The Aerospace Corporation, 14301 Sullyfield Circle, Unit C, Chantilly, VA 20151-1622 USA}

\author{Dana I. Casetti-Dinescu}
\affiliation{Department of Physics, Southern Connecticut State University, 501 Crescent Street, New Haven, CT 06515, USA}

%\collaboration{(AAS Journals Data Scientists collaboration)}



\begin{abstract}
The \textit{Kepler} mission and subsequent ground-based follow-up observations have revealed a 
number of exoplanet host stars with nearby stellar companions. This study presents speckle observations 
of 57 {\it Kepler} objects of interest (KOIs) that are also double stars, each observed over a 3 to 8 year period, which has allowed us to track their relative motions with high precision. 
Measuring the position angle 
and separation of the companion with respect to the primary can help determine if the pair exhibits 
common proper motion, indicating it is likely to be a bound binary system. We report on the motions of 34 
KOIs that have close stellar companions, three of which are triple stars, for a total of 37 companions studied. 
Eighteen of the 34 systems are confirmed exoplanet hosts,
including one triple star, 
while four other systems have been subsequently judged to be false positives and twelve are yet to be confirmed as planet hosts. 
We find that 21 are most likely to be common proper motion pairs, 4 are line-of-sight companions, and 12 are 
of uncertain disposition at present. 
The fraction of the confirmed exoplanet host systems 
that are common proper motion pairs is approximately 86\% in this sample. In this subsample, 
the planets are exclusively found with periods of less than 110 days, so that in all cases the stellar 
companion is found at a much larger separation from the planet host star than the planet itself. A 
preliminary period-radius relation for the confirmed planets in our sample suggests no obvious differences at this stage with the full sample of known exoplanets.

%The exoplanets are in tight orbits ($<$40 days) while the binary stars tend to have very long projected orbital periods, over a thousand years for most. However, four stellar pairs orbit each other closer than 70 AU. Two binary systems show preliminary evidence of possible coplaner stellar and planetary orbits.
%This on-going work, and its extension to the far closer K2 and TESS exoplanet host binaries, is a first observational study of the properties and architectures of binary stars which harbor exoplanets.
\end{abstract}


\section{Introduction}
NASA's \textit{Kepler} mission, launched in 2009, monitored the brightness of approximately 170,000
stars in a wide field of view, and employed the transit technique to identify the presence of other celestial 
bodies orbiting the host star.  Over 2600 of these detections have since been confirmed as signatures of 
transiting exoplanets. A number of studies have focused on ground-based follow-up and 
analysis of {\it Kepler} data, for example, \citet{2011ApJ...742...38Y};
%\citet{2014ApJ...295...60H};  
\citet{eve15}, \citet{tor17}, \citet{2017AJ...153...117H}, \citet{2017AJ...153...2F}, \citet{may18}. 
%\citet{how19}. 
In general, roughly half of all Sun-like stars in the nearby field population are part of multi-star systems 
\citep{2010ApJS...190...1R}, and 
high resolution speckle imaging observations of {\it Kepler} and {\it K2} exoplanet host stars have shown that 
the multiplicity fraction may also be nearly as high as the field 
\citep{2014ApJ...295...60H, 2018AJ....156...31M}, at least over the range of 
projected separations applicable in those studies. Other studies have presented 
evidence that the multiplicity rate of exoplanet host stars differs from the field population depending on
separation, and that exoplanet hosts tend not to have stellar companions at separations of 10's to 100's
of AU \citep{2014ApJ...791..111W, 2015ApJ...806..248W, 2015ApJ...813..130W, kra16, zie20}.

Identifying and characterizing these binaries throughout the full range of bound separations is important for two 
principal reasons. First, the existence of stellar companions can pose a problem to planetary radii calculations 
from transit curves due to transit depth dilution from the companion. If a companion falls within the 
{\textit{Kepler} aperture and the system is assumed to be a single star, the planetary radii will be underestimated 
on average by a factor of 1.5 \citep{2015ApJ...805...16C}. Systematic errors can then also occur for the 
planets' derived atmospheric 
structure and mean density \citep{2017AJ....154...66F}, and even in the derived properties of the 
host star itself \citep{fur20}. Second, one of the major goals of exoplanet 
science is the determination of the occurrence rate of different types of planets and the conditions 
under which they form and evolve. The study of binaries where one or both stars is an exoplanet host can 
serve as a tool to address these issues by providing robust statistics of stellar properties and orbital 
characteristics that 
can be used to compare to star and planet formation theories; for a recent review see e.g.\ 
\citet{2020SSRv..216...70L}. 

 %Observationally, direct determination 
%of the orbit of the binary pair, leading to mass ratios and robust tests of star and planet 
%formation, will be an effective method that we can now begin to employ. Characterizing the binary companions and making high-resolution observations over time allows for the determination of the orbit characteristics of the binary system, such as the inclination angle of the binary orbit relative to that of the planet or planets. Furthermore, statistical studies of the correlation between binary and planet parameters will inform star and planet formation models. 

However, once a companion star is detected, it may be either a line-of-sight companion (forming an optical double with the primary) or a gravitationally 
bound component (forming a binary system with the primary). Previous work by \citet{eve15} 
%\citet{2013ApJ...771...107H} 
and \citet{2017AJ...153...117H} used photometric information derived from high-resolution speckle and 
adaptive optics imaging to place the components of {\it Kepler} double stars on the H-R diagram. This allowed 
for a characterization of each system as bound or unbound based on whether the two stars fell on a common 
isochrone, where both stars were assumed to be at the same distance and in most cases both stars 
would be found near the main sequence. There was of course an inherent degeneracy
between whether the secondary star was a companion dwarf or a background giant, and in addition, the photometric uncertainties were too large in some cases to provide a definitive assessment.

This paper reports on a different, complementary approach. We have completed a long-term observational 
program using speckle observations of an initial sample of over 50 well-characterized \textit{Kepler} 
double stars in order to determine whether the stars are gravitationally bound via astrometric analysis. Given 
the typical distances for \textit{Kepler} host stars, roughly $\sim$200 pc to 1.4 kpc in general, the separations of 
the primary and secondary stars that can be detected are on the order of hundreds to thousands of AU. This 
means that for any bound companions, the orbital periods will generally be very long -- hundreds or thousands
of years -- and their positions relative to one another will change very slowly. However, it can still often be
shown that these systems are likely to be a gravitationally bound pair by tracking their common proper motion 
on the sky over a much shorter time baseline than their orbital period. Given the astrometric precision possible 
with speckle imaging,
observations over as short as a few years are sufficient. These relative astrometry data, in combination with 
system proper motions from {\it Gaia} \citep{gai18}, are used to make a final assessment. In this paper, we 
present the main body of our observational results, and we attempt to characterize each system as likely to be 
a common proper motion pair or a line of sight companion.

%Four of the binary pairs we analyze here are now known to be False Positives. They were added to our watch list early on in the \textit{Kepler} mission and as such, we have astrometric measurements for them. We call them out as FPs in the text but include their analysis here anyway.

\section{Observations and Data Reduction Methodology}

%This study is focussed on the motions of stellar companions to KOIs as derived from an astrometric 
%analysis of diffraction-limited imaging data using speckle imaging. 
Two different speckle instruments were used for work on this project.
 The majority of the observations were taken
with the Differential Speckle Survey Instrument (DSSI), which was completed in 2008 at Southern Connecticut
State University \citep{2009AJ...137...5057H} and subsequently became a visitor instrument at the 
WIYN Telescope. In early 2010, the two low-noise, large-format CCDs originally used to record the 
speckle images were upgraded to two Andor iXon 897 EMCCD cameras. 
This change made it possible to take diffraction-limited data on much fainter stars than before; the limiting magnitude of DSSI at WIYN with the CCDs was $V=12.5$ in good conditions, 
whereas with the EMCCDs it was approximately 14.5. The instrument was then subsequently used in 
follow-up work for the {\it Kepler} mission over the next few years \citep{how11}, 
in addition to performing a large 
%NSF-funded 
survey of nearby binary stars at WIYN \citep{2011AJ...141...45H} and a number of other smaller projects. 
Hundreds of Kepler Objects of Interest (KOIs) were observed, 
and dozens were found to have stellar companions with separations less than 2 arc seconds. DSSI 
was also used on the Gemini-North
Telescope (2012-2016) and the Lowell Discovery Telescope (LDT, 2014-present) for the same survey 
projects.

In 2016, a successor instrument to DSSI, the NNExplore Exoplanet Stellar Speckle Imager
(NESSI) \citep{2018PASP..130e4502S}, was completed and began operations at WIYN. In the 2016B and 2017A observing semesters, NESSI was used to obtain further observations of KOIs known to have close stellar 
companions. Together with the earlier observations of the same stars, the full data set represents a unique 
opportunity to study the proper motions of the companions relative to their primary stars over a longer 
time baseline. In combination with data from the {\it Gaia} satellite, a full picture of the motions of the 
companions can be drawn.
% that is the goal of this paper. Once complete, statements regarding the likely 
%physical association of the companion to the primary can be made, which is an important step in determination 
%of the environment of exoplanets in the system, if they are confirmed.

The DSSI filters had center wavelengths of 692 and 880 nm, with a FWHM  of 40 and 50 nm respectively; for NESSI observations, the filters were centered at 562 and 832 nm, with FWHM transmission of 44 and 40 nm, respectively. For the stars discussed in this paper, a minimum of 1000 speckle frames per observations was generally taken for brighter stars, and up to $\sim$15,000 frames for the faintest targets. Typical frame integration times of 40 ms were used at WIYN and the LDT, and 60 ms at Gemini. The total number of frames taken for a given observation depended on a combination of target brightness and observing conditions. The data files were stored as FITS cubes in 1000-frame blocks and co-added after the fact, if more than one block was taken.

To reduce the data, we used the general methodology described in previous papers, such as 
\citet{1996AJ...111...1681H}, \citet{2009AJ...137...5057H}, and \citet{2011AJ...141...45H}. 
The average autocorrelation of the speckle frames for a given observation and filter are computed and Fourier 
transformed to arrive at the spatial frequency power spectrum of the observation. The same is done for a 
``calibration'' point source, i.e. an unresolved single star observed near on the sky and near in time to the 
science target.
The average triple correlation is also formed for the science target and Fourier transformed; this results in the image bispectrum \citep{1983ApOpt..22.4028L}. The Fourier transform of the object is then formed by first obtaining its modulus and combining that with an estimate of the phase. The former is arrived at by simply dividing the power spectrum of the object by that of the calibration point source and taking the square-root, and the latter is derived from the bispectrum using a relaxation algorithm first developed by \citet{1990JOSAA...7...001243}. Once the Fourier transform of the object is in hand, a reconstructed image is formed by low-pass filtering (to mitigate high frequency noise) and inverse-transforming.  

The reconstructed images are then used as a starting point to identify the location of any detected companion stars in order to begin the process of obtaining astrometry for them. The initial rough astrometry, obtained as 
the pixel location of the maximum of the secondary peak in the reconstructed image, is then used as input for a weighted least-squared fitting program for the object power spectrum first described in \citet{1996AJ...111...1681H}. This program uses Fourier-space results to output the final relative photometry and astrometry for each observation. 
An important key to getting well-calibrated astrometry is to have a precise value for the pixel scale and image orientation for each observation; this will be discussed in Section 4. 
%We gather this information using both calibration images of a slit mask and a regimen of observing ``calibration binaries", systems which have extremely well-determined stellar orbits (usually from long baseline optical interferometry observations). 
%While obtaining high-precision astrometric calibrations over a period of years and at different telescopes is 
%challenging, the effort and 
%care that we have made in maximizing the long-term astrometric performance has provided the 
%basis for the current study. As discussed more fully in e.g.\
%\citet{2011AJ...141...45H} (for WIYN data), \citet{2012AJ....144..165H} (for Gemini-N), and 
%\citet{2015AJ....150..151H} (for the LDT),
%our astrometric results during the period of observations presented here generally have 1-$\sigma$ un
%certainties of between 1 and 3 mas in position in all cases.

%While no similarly detailed study of a large data set obtained with NESSI exists at present, it is 
%probably reasonable to expect that it has similar properties in terms of its detection limit and astrometric
%precision, given NESSI's similarity in design and performance to DSSI. The initial indications from \citet{2018PASP..130e4502S} suggest this as well.

\section{Results}

In total, observations of 57 KOI stars with at least one known stellar companion were obtained 
over a time frame of 
approximately eight years using a combination of the three telescopes and two instruments 
mentioned above. Thirty-four of these objects were
observed during at least three different epochs, and of these, three were seen as triple stars
in our observations. 

Table 1 presents the final differential astrometry and photometry from the speckle observations. The columns 
give (1) the KOI number; (2) the Kepler number in
cases where the existence of one or more exoplanets in the system has been subsequently confirmed; (3)
the right ascension and declination in Washington Double Star (WDS) Catalog format; (4) the Besselian year
of the observation; (5) the position angle of the secondary star, with north through east defining the
positive sense of the angle; (6) the separation of the two stars; (7) the magnitude difference of the pair;
(8) the center wavelength of the filter used; (9) the full-width at half-maximum for the transmission of the
filter; and (10) the telescope and instrument combination used. Generally, two observations are listed for
each epoch in Table 1 because both DSSI and NESSI record speckle patterns in two colors
simultaneously.
However, there are occasional cases where the observation in one of the filters was judged to be too low
in quality to produce a reliable result; a typical case would be when the secondary is faint and red so that
the magnitude difference is lower in the redder filter than in the bluer filter of the two observations. ``Unpaired''
observations are reported in Table 1 in such cases. All of the companions listed in Table 1 have been
previously noted by other authors, such as \citet{hes18}, \citet{2017AJ...153...2F}, \citet{2017AJ...153...117H},  \citet{eve15}, \citet{2014ApJ...295...60H}, \citet{2012AJ....144..165H}, and \citet{how11};
however, the measures we show here
supersede any for the same epochs in those papers. This is because the data reduction used 
in the current analysis more
rigorously defines the astrometric calibration and precision of the data set, 
and also because care has been taken here to note
when the magnitude differences are affected by speckle decorrelation, as described in e.g.\  \citet{hor17}.
In addition, Table 1 contains later epochs of observation that did not appear in any of the above references.

To characterize the data set as a whole, we plot in Figure 1 the magnitude difference of measures from Table 1
as a function of both $V$ magnitude of the source and the 
log of the separation of the companion. In Figure 1(a), the
magnitude distribution is seen to be mainly between 10th and 14th magnitude, considerably fainter than the
other large binary star speckle programs at WIYN with DSSI; \citet{hor17} is one example of a comparably
large data set on brighter stars. In previous papers, including \citet{hor17} and references therein,
the detection limit of the DSSI instrument at WIYN is also estimated. In Figure 1(b), in addition
 to the data from Table 1, we show the detection limit curve for DSSI from that work as a dashed line, 
 and it can be
seen that the measures in Table 1 span most of discovery space of the instrument; however, in the range
of separations from roughly 0.1 to 0.3 arc seconds there is a noticeable lack of large magnitude-difference
companions, and throughout the plot there are few systems that are close to the limiting magnitude curve.
This may suggest that, for the range of magnitudes represented here, the typical detection limit is lower
than the curve shown by as much as several tenths of a magnitude. 
This is not unexpected, due to the lower signal-to-noise data obtained for these fainter sources relative to 
other survey work with the same instrument. For this reason, we averaged the standard DSSI detection
curve with that of the faint point source observed at WIYN that appears in \citet{2014ApJ...295...60H}, 
and that is shown as the solid blue curve in Figure 1(b).
We will assume that this is a better representation of our overall detection capabilities for the purposes
of the analysis here.

\begin{deluxetable}{rllrlrrrrl}
\tabletypesize{\scriptsize}
\tablewidth{0pt}
\tablenum{1}
\tablecaption{KOI Double Star Speckle Measures}
\tablehead{
\colhead{KOI} &
\colhead{Exoplanet} &
\colhead{WDS} &
\colhead{Date} &
\colhead{$\theta$} & \colhead{$\rho$} &
\colhead{$\Delta m$} &
\colhead{$\lambda$} &
\colhead{$\Delta \lambda$} &
\colhead{Tel., Inst.} \\
\colhead{Number} &
\colhead{System} &
\colhead{($\alpha$,$\delta$ J2000.0)} &
\colhead{(2000+)} &
\colhead{($^{\circ}$)} & \colhead{(${\prime \prime}$)} &
\colhead{(mag)} &
\colhead{(nm)} &
\colhead{(nm)} &
\colhead{and Notes\tablenotemark{a}}}

\startdata
1 & Kepler-1 & $19072+4919$  & 11.4480  & 136.7 & 1.1060 & $<$ 4.75 & 692 & 40 & WD, b \\
1 & Kepler-1 & $19072+4919$  & 11.4480  & 136.0 & 1.1055 & $<$ 3.49 & 880 & 50 & WD, b \\
1 & Kepler-1 & $19072+4919$  & 13.7227  & 135.0 & 1.1016 & $<$ 4.40 & 692 & 40 & WD, b \\
1 & Kepler-1 & $19072+4919$  & 13.7227  & 136.5 & 1.1024 & $<$ 3.26 & 880 & 40 & WD. b \\
1 & Kepler-1 & $19072+4919$  & 13.7284  & 136.4 & 1.1107 & $<$ 4.25 & 692 & 40 & WD. b \\
1 & Kepler-1 & $19072+4919$  & 13.7284  & 136.3 & 1.1093 & $<$ 3.41 & 880 & 40 & WD, b \\
1 & Kepler-1 & $19072+4919$  & 17.2747  & 136.2 & 1.1101 & $<$ 3.45 & 832 & 40 & WN, b \\
1 & Kepler-1 & $19072+4919$  & 17.2747  & 136.2 & 1.1110 & $<$ 4.95 & 562 & 40 & WN, b \\
13 & Kepler-13 & $19079+4652$ & 10.4649 & 279.7 & 1.1647 & $<$ 0.84 & 692 & 40 & WD, b \\
13 & Kepler-13 & $19079+4652$ & 10.4732 & 279.8 & 1.1619 & $<$ 0.90 & 692 & 40 & WD, b \\
%13 & Kepler-13 & $19079+4652$ & 10.4732 & 279.6 & 1.1667 & $<$ 1.20 & 562 & 40 & WD, b \\
\enddata

\tablenotetext{a}{Telescope and Instrument Combinations are abbreviated as follows: WD = WIYN+DSSI;
LD = LDT+DSSI, GD = Gemini-North+DSSI; WN = WIYN+NESSI.}
\tablenotetext{b}{ The magnitude difference appears here as an upper limit due to the speckle decorrelation
effect discussed in the text.}

\tablecomments{Table 1 is published in its entirety in the machine-readable format.
      A portion is shown here for guidance regarding its form and content.}

\end{deluxetable}

\begin{figure}[!t]
\figurenum{1}
\plottwo{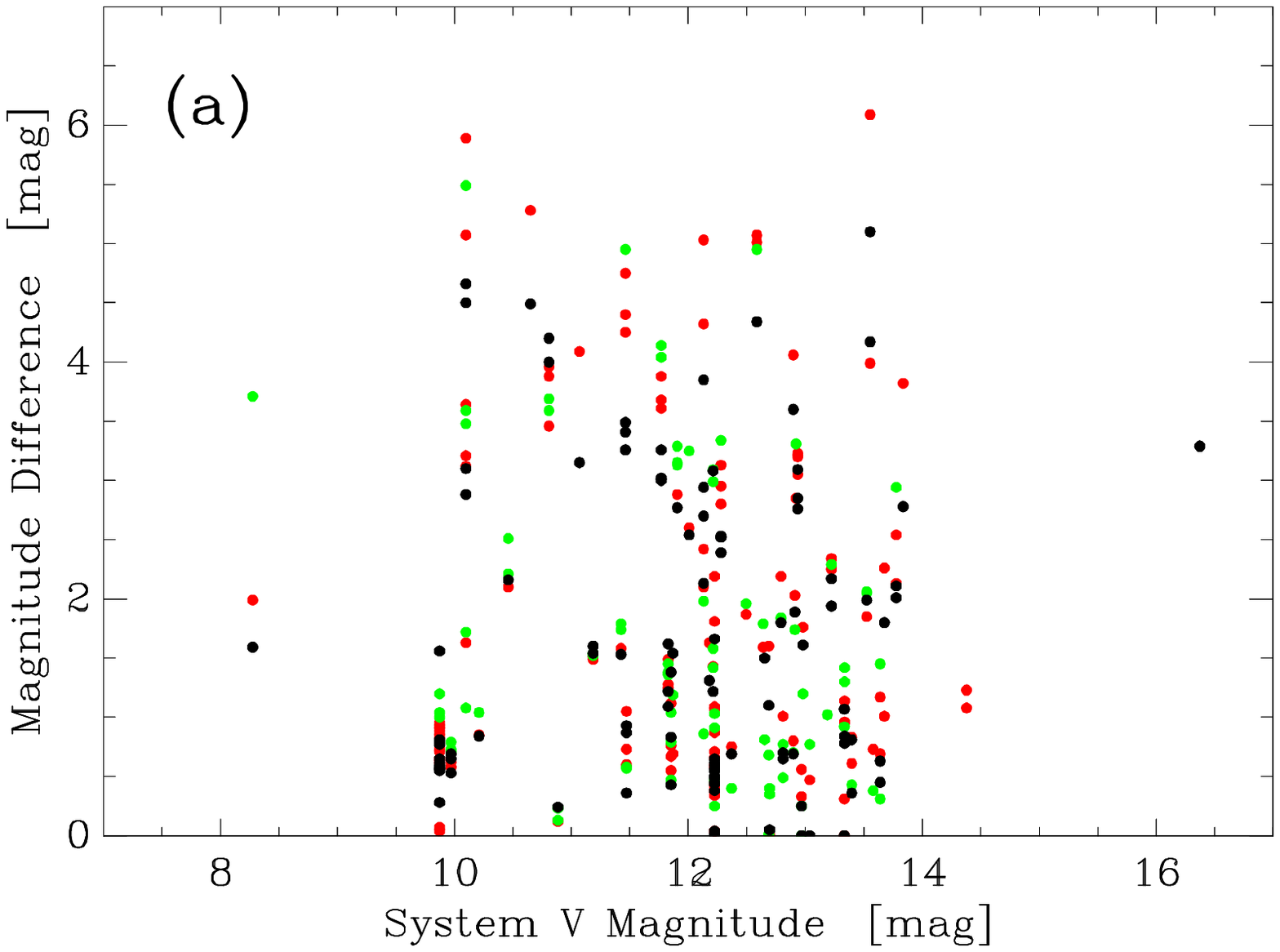}{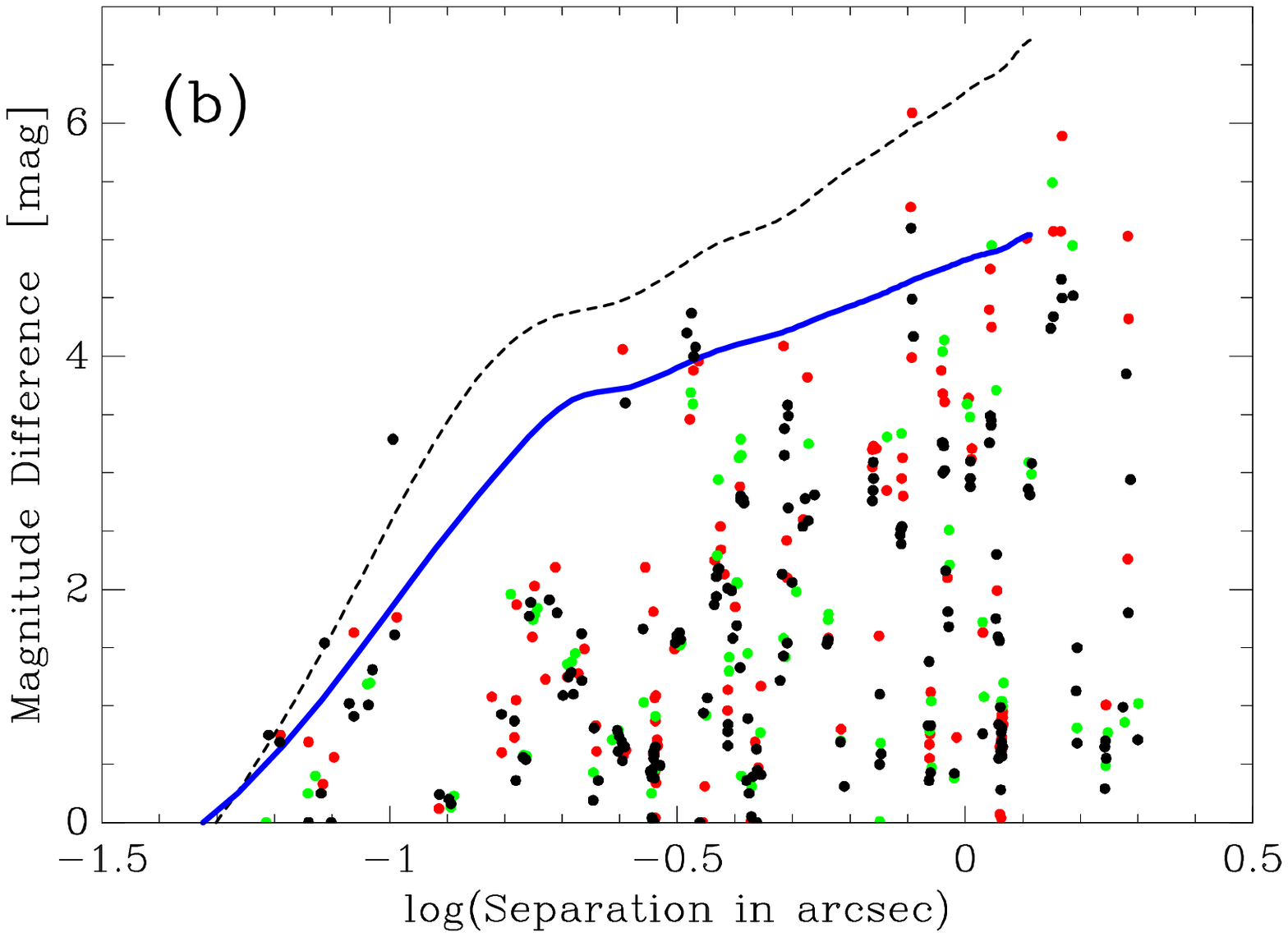}
\figcaption{(a) Magnitude difference for all measures listed in Table 1 as a function of the 
system V magnitude appearing in the literature. 
(b) The same values plotted versus the log of the separation appearing in Table 1.  The
dashed curve is the detection limit curve for DSSI at the WIYN at 692 nm, taken from \citet{hor17},
which as discussed in the text, may not represent the data here as well due to the faintness of
these stars. The blue curve is the result of averaging the standard curve with that of a fainter source,
which appears to represent the detection limits here better on average.
In both plots, the color of the plot symbol indicates the wavelength of the observation:
green is used for 562 nm,
red is used for 692 nm, and black is used for both 832 and 880 nm.}
\end{figure}

\section{Astrometric Precision}

Since our principal aim is to establish if the companions in Table 1 are likely to be gravitationally bound to their
primary stars, it is important to characterize the precision of the astrometry we obtain.
 This in turn is dependent
on the derivation of the pixel scale and detector orientation relative to celestial coordinates. For data
taken at WIYN with DSSI, a well-established regimen of scale calibration was in place for the entire tenure
of the instrument at that site. It was based on the use of a slit mask mounted to the telescope's
tertiary mirror baffle
support structure, in the converging beam of the telescope. The distance between slits was well-measured,
as was the distance from the mountings surface to the telescope focal plane. When observing a very bright
unresolved star with the mask in place, a series of fringes of different spacings is produced in the image of the
star on the detector. The effective wavelength of the observation was determined using a spectrum of a
star with the same type chosen from the spectral library of \citet{pic98} and combining that flux
curve with the known transmission curve of the filter used and the quantum efficiency curve of the detector.
In combination with the distance measures mentioned just above and the focal length of the telescope,
the fringe spacing can then be determined in arc seconds. By measuring the number of pixels across the
fringe pattern, the scale in arc seconds per pixel is obtained. However, the actual measurement is done in the
Fourier domain, where the fringes map to sharp line-like features at particular spatial frequencies. A least-squares
fit is then performed between the data and a model function representing the pattern of lines where the scale
value is the fitted parameter.

The orientation of the detector relative to celestial coordinates was obtained by recording a set of offset images
of bright stars, generally once per night. An initial image is taken, and then small offsets in position in different
cardinal directions are made with the telescope. The images are reduced and the position of the star in each
image is determined via centroiding. This yields the offset angle the orientation of the detector pixels and 
the cardinal directions north and east on the sky. It also
yields an estimate of the scale, but it is generally lower precision than the measures obtained 
from the slit mask.
In some runs, if offset sequences were judged to be of lower quality (for example, if the images were taken in
windy conditions or in poor seeing), then the measured orientation angle is supplemented with the value inferred
from a small number of binaries observed for the purpose of monitoring scale and orientation over time
and referred to as ``scale'' binaries; these systems
generally have extremely high-precision orbits determined from long-baseline optical interferometry (LBOI).

At the LDT and at Gemini, no slit mask was available, so a regimen of observing scale binaries from LBOI 
was employed, and both scale and orientation were derived solely from those measures. With NESSI 
observations
at WIYN, the slit mask was of course available, but to avoid removing and reinstalling the tertiary
mirror baffle for each speckle observing run, the mask was instead mounted to the baffle itself. Since that change,
the slit mask scale determinations have been less consistent from run to run compared with the 
DSSI scale values,
and so for the work presented here, we fell back to the same regimen of ``scale'' binaries as used at the LDT
and Gemini when reducing NESSI data. Overall, despite these differences in approach, the scale values
appear to be accurate to $\sim$0.2\% or better, and the orientation angles to $\sim$0.3$^{\circ}$, when
judged against binaries with well-determined orbits (that were not used in scale determinations).

While obtaining high-precision astrometric calibrations over a period of years at different telescopes is 
challenging, the effort 
that we have made to maximizing the long-term astrometric performance provides the 
basis for the current study. As discussed more fully in e.g.\
\citet{2011AJ...141...45H} (for WIYN data), \citet{2012AJ....144..165H} (for Gemini-N), and 
\citet{2015AJ....150..151H} (for the LDT),
the astrometric results during the period of observations presented here generally have 1-$\sigma$ uncertainties of between 1 and 3 mas in position in all cases for the standard program of binary star
survey observations.

\begin{figure}[!t]
\figurenum{2}
\plottwo{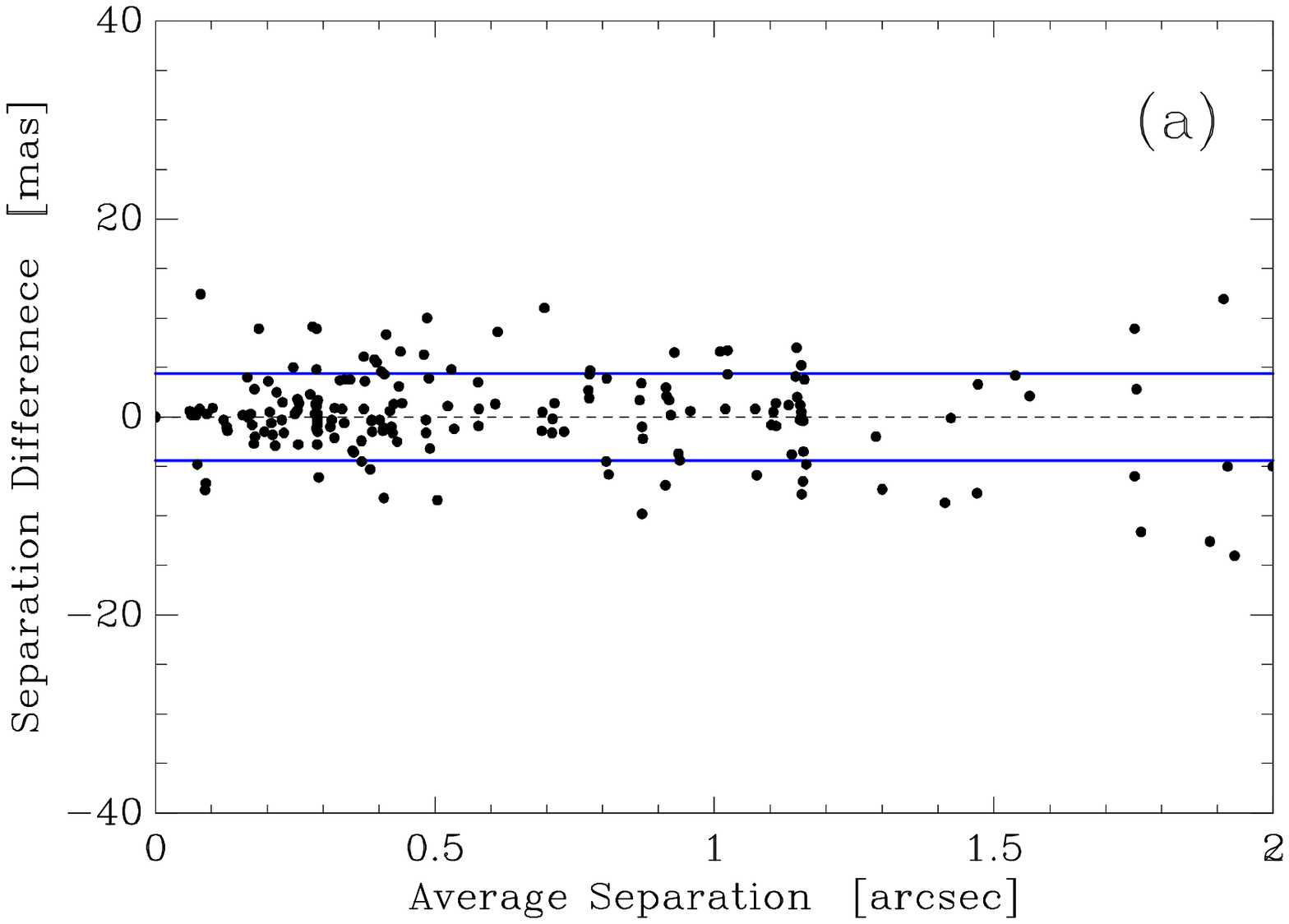}{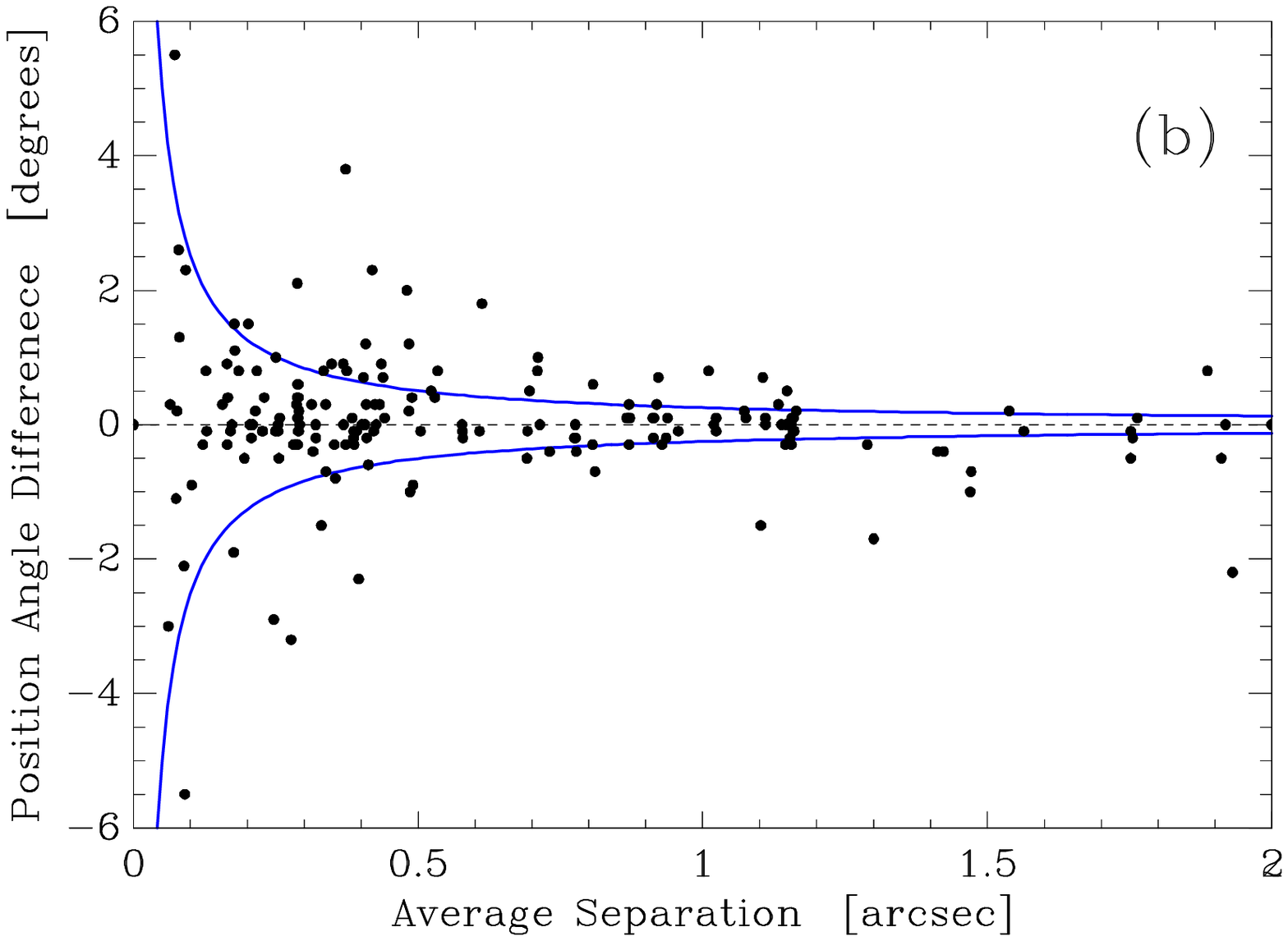}
\figcaption{(a) Difference in the separation as a function of average separation for the paired measures in Table 1. The zero line is drawn to guide the eye, and the blue dashed lines represent $\pm4.4$ mas, the standard deviation
of these values.
(b) Differences in position angle as a function of average separation. Again, the zero line is drawn, and in this
case the blue dashed curves represent the $\pm$1-$\sigma$ uncertainty implied by a linear uncertainty of
4.4 mas.}
\end{figure}

A unique capability of the DSSI and NESSI instruments is their ability to record two speckle
patterns in two filters simultaneously. For astrometric studies, this provides a natural way to characterize the
internal astrometric precision of a data set, simply by comparing results obtained independently from the two
channels of the instrument, which should yield the same result. In Figure 2(a), we show
the separation differences as a function of average separation when subtracting the result in one 
channel versus
the other for all the paired observations appearing in Table 1. Figure 2(b) shows a similar plot for position angle
residuals. Separation residuals have an average value of $0.26 \pm 0.32$ mas, with a standard deviation of
$4.38 \pm 0.23$ mas. There are no obvious trends in the plot, except perhaps a slight increase in the scatter at
the largest separations. It is in this region of the plot that any systematic errors in the determination of the scale 
will be
seen in addition to the random scatter of the measures; if the scale is slightly overestimated for some runs and
slightly underestimated for others, this could lead to a broadening of the residual distribution that is most
noticeable at the largest separations. However, there are few data points in this region of the plot which at worst have a modest increase in the scatter, and so
we conclude that this is at most a small contribution to the astrometric uncertainty of the data set overall.

In to Figure 2(b), the average value obtained is again close to zero, $0.05 \pm 0.08^{\circ}$, and
the position angle residuals show the typical increase in standard deviation at small separations. (The
average values for the standard deviation throughout the plot is $1.03 \pm 0.05^{\circ}$.) This increase is
a geometrical effect; if the astrometric uncertainty is equal and independent in the direction
orthogonal to the separation, then the position angle uncertainty (in radians) is given by

 \beq
 \delta \theta = \frac{\delta \rho}{\rho},
 \eeq

\noindent
where $\delta \theta$ and $\delta \rho$ are the uncertainties in position angle and separation respectively 
$\rho$ is the separation of the pair.
In Figure 2, we have placed curves on the plot using $\delta \rho = \pm 4.4$ mas, and we see that, in the mean,
this appears to provide a reasonable 1-$\sigma$ envelope to the data at all separations. A few measurements
have relatively large differences in position angle at large separations, but many of these are extremely faint
companions, and harder to fit on that basis.

The differences plotted in Figure 2 are the subtraction of two independent measurements with presumably
the same uncertainty, in which case the subtraction has an uncertainty that is $\sqrt{2}$ larger than the
uncertainty of either individual measure. This means that the uncertainty in a single measure in Table 1
is given by $4.38 \pm 0.23$ divided by $\sqrt{2}$, or in other words, $3.10 \pm 0.16$ mas. This is larger 
than the measurement uncertainty derived in e.g. \citet{hor17} for DSSI at WIYN; in that paper the value 
stated was $1.73 \pm 0.04$ mas. The main difference between that data set and the one in Table 1 is the magnitude range of the
target stars: in \citet{hor17}, the stars were mainly in the range $6 < V < 10$, whereas for the current
sample it is $10 < V < 14$. Therefore the deterioration in the astrometric precision is most likely due to
lower signal-to-noise ratios for the stars in this study. Given the consistency between the separation and
position angle differences, we will assume moving forward that the linear measurement precision of single
measures in Table 1 is 3.1 mas. Likewise, we assume that the average uncertainty in position angle for
measures in Table 1 is $1.03^{\circ} / \sqrt{2} = 0.7^{\circ}$.

\section{Relative Proper Motion Determination of the Secondary with Respect to the Primary Star}

\begin{deluxetable}{rrrrrrll}
\tabletypesize{\tiny}
\tablewidth{0pt}
\tablenum{2}
\tablecaption{Proper Motions for KOIs Observed in at Least Three Epochs}
\tablehead{
\colhead{KOI} &
\colhead{Kepler No.} &
\colhead{Speckle $\Delta \mu_{\alpha}$} & 
\colhead{Speckle $\Delta \mu_{\delta}$} &
\colhead{System $\mu_{\alpha}$} & 
\colhead{System $\mu_{\delta}$} &
%\colhead{Distance} &
\colhead{$\pi$} &
\colhead{Source} \\
\colhead{No.} &
\colhead{or Disp.\tablenotemark{a}} & 
\colhead{(mas/yr)} &
\colhead{(mas/yr)} &
\colhead{(mas/yr)} &
\colhead{(mas/yr)} &
%\colhead{(pc)} &
\colhead{(mas)} &
\colhead{for $\mu$, $\pi$\tablenotemark{b}}
}
\startdata
1 & 1 & $0.852 \pm 1.124$ & $-0.417 \pm 1.540$ & $5.219 \pm 0.043$ & $1.619 \pm 0.043$ & 
$4.615 \pm 0.022$ & DR2 \\
13 & 13 & $1.042 \pm 0.545$ & $0.272 \pm 0.554$ & $-4.401 \pm 0.187$ & $-15.780 \pm 0.237$ &
$1.905 \pm 0.105$ & DR2\tablenotemark{c} \\
98 &  14 & $-0.793 \pm  0.328$ & $0.277 \pm 0.237$ & $1.0 \pm 1.3$ & $-10.2 \pm 1.0$ &
$1.449 \pm 0.113$ & UCAC4, CFOP \\
118 & 467 & $-16.526 \pm 0.932$ &  $-42.943 \pm 0.645$ & $14.227 \pm 0.038$ & $35.604 \pm 0.041$ &
$2.116 \pm 0.023$ & DR2\tablenotemark{c} \\
120 & (PC) & $-6.503 \pm 8.617$ & $-5.921 \pm 5.629$ & $-6.9474 \pm 0.094$ & $-12.249 \pm 0.097$ &
$1.240 \pm 0.044$ & DR2\tablenotemark{c} \\
177 & (PC) & $0.716 \pm 0.118$ & $0.238 \pm 0.161$ & $5.5 \pm 3.5$ & $6.1 \pm 1.4$ &
$2.160 \pm 0.155$ & UCAC4, CFOP\\
258AB & (FP) & $1.089 \pm1.118$ & $1.012 \pm 0.675$ & $4.183 \pm 0.052$ & $-7.309 \pm 0.055$ &
$2.648 \pm 0.031$ & DR2 \\
258AC & (FP) & $-15.745 \pm1.597$ & $-4.521 \pm 3.283$ & $4.183 \pm 0.052$ & $-7.309 \pm 0.055$ &
$2.648 \pm 0.031$ & DR2 \\ %remove?
%270 & 449 & $2.431 \pm 0.189$  & $1.166 \pm 0.234$ &  \\
%279 & 450 & $0.395 \pm 0.693$ & $-1.205 \pm 0.799$ &  \\
%284 & 132 & $-0.225 \pm 0.469$ & $-0.826 \pm 0.397$  & \\
%307 & 520 & $2.464 \pm 0.918$ & $0.655 \pm 0.587$ & \\
%640 & 632 & $-1.519 \pm 0.765$ & $1.717 \pm 0.970$ \\
%959 &  & $3.476 \pm 2.104$ & $1.238 \pm 3.191$ & \\
%976 &  & $-0.242 \pm 0.537$ & $1.918 \pm 0.209$ & \\
%977 &  & $0.165 \pm 0.600$ & $-0.120 \pm 0.662$ & \\
%980 &   & $1.689 \pm 0.298$ & $0.736 \pm 0.657$ & \\
%984 &  & $0.898 \pm 0.508$ &  $-1.517 \pm 1.354$ & \\
%1119 &  & $4.152 \pm 2.817$ & $-0.461 \pm 3.355$ & \\
%1150 & 780 &  $-2.033 \pm 0.922$ & $0.006 \pm 0.472$ & \\
%1531 & 884 & $-0.184 \pm 0.834$ & $-0.585 \pm 1.297$ & \\
%1613 & 907 & $0.670 \pm 0.311$ & $1.168 \pm 0.693$ & \\
%1792AB & 953 & $-17.895 \pm1.135$ & $-32.149 \pm 0.698$ \\
%1792AC & 953 & $-19.662 \pm 3.302$ & $-28.702 \pm 5.376$ & \\
%1890 & 1002 & $1.929 \pm 1.067$ & $0.679 \pm 0.644$ & \\
%1962 &  & $2.216 \pm 0.267$ & $-1.046 \pm 0.199$ & \\
%2059 &  & $-1.496 \pm 0.653$ & $-1.189 \pm 0.406$ & \\
%2754 & 1339 & $0.493 \pm 0.399$ & $0.094 \pm 0.693$ & \\
%2837 & 1364 & $2.408 \pm 0.940$ & $1.421 \pm 0.929$ \\
%2904 & 1382 & $-14.982 \pm 3.276$ &$2.458 \pm 0.910$ & orbit? \\
%3020 &  & $1.208 \pm1.469$ & $0.532 5\pm 2.955$ & remove? \\
%3156 &  & $-2.537 \pm 3.608$ & $7.012 \pm 2.560$ & remove? \\
%3214AB &  & $-1.255 \pm 0.874$ & $1.523 \pm 1.528$ & \\
%3214AC &  & $0.816 \pm 5.314$ & $3.787 \pm 1.704$ & \\
%5578 &  & $2.279 \pm 0.305$ & $0.257 \pm 0.291$ & \\
%5822 &  & $-0.255 \pm 0.582$ & $-2.281 \pm 2.640$ &  \\
\enddata

\tablenotetext{a}{If no Kepler number is given, the disposition as either a planetary candidate
        (PC) or false positive (FP) is given, based on information available on the Kepler 
         Community Follow-up Observing Program (CFOP) website, {\tt https://exofop.ipac.caltech.edu/cfop.php}.}
\tablenotetext{b}{Abbreviations are as follows:
CFOP = Kepler Community Follow-up Program website,
DR2 = \citet{gai18}, 
Tycho-2 = \citet{hog00},
UCAC4 = \citet{zac13}.}
\tablenotetext{c}{The pair is resolved in DR2, so the values shown are for the primary star.}

\tablecomments{Table 2 is published in its entirety in the machine-readable format.
      A portion is shown here for guidance regarding its form and content.}

\end{deluxetable}

\begin{figure}[!t]
\figurenum{3}
\plottwo{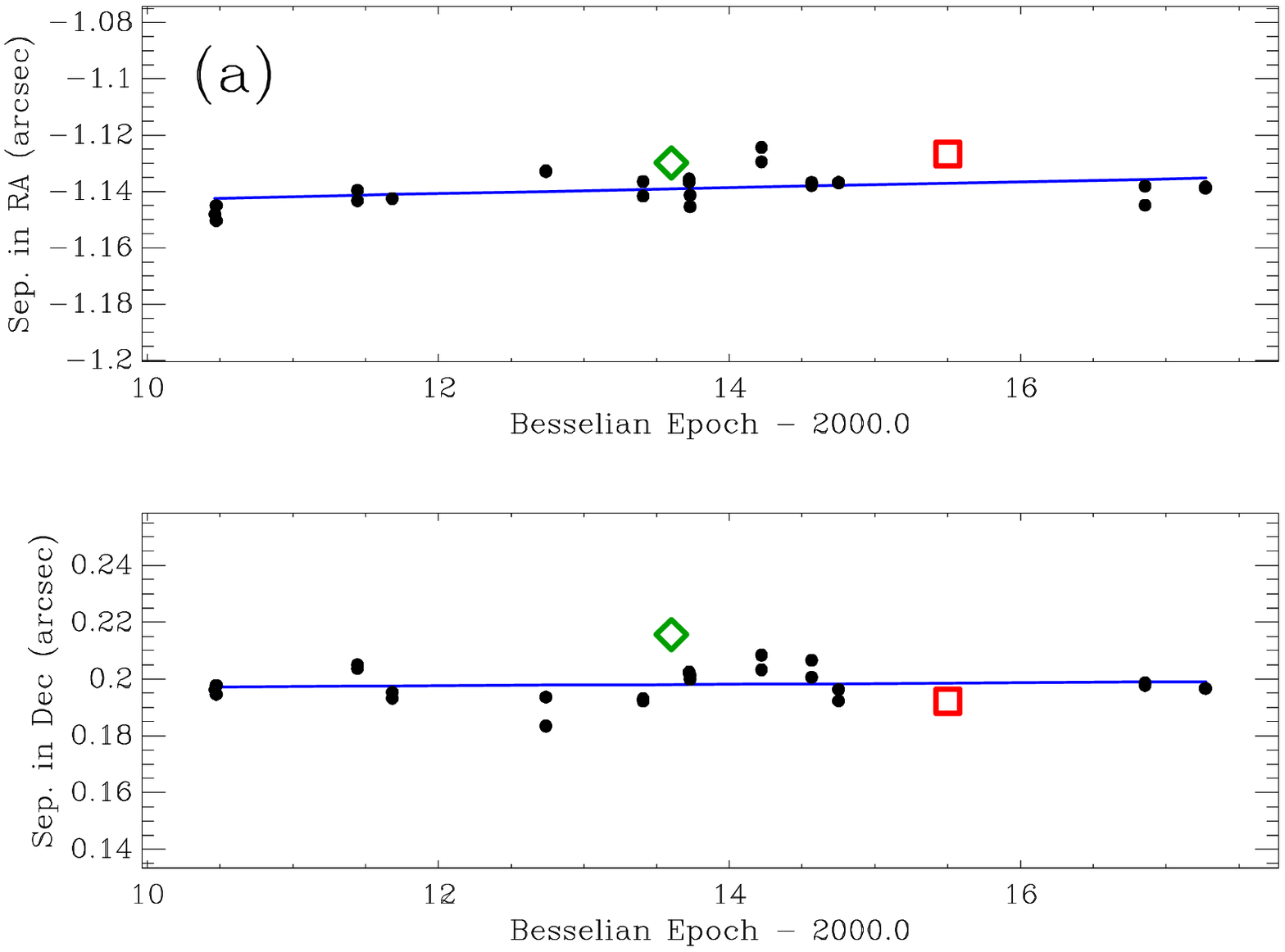}{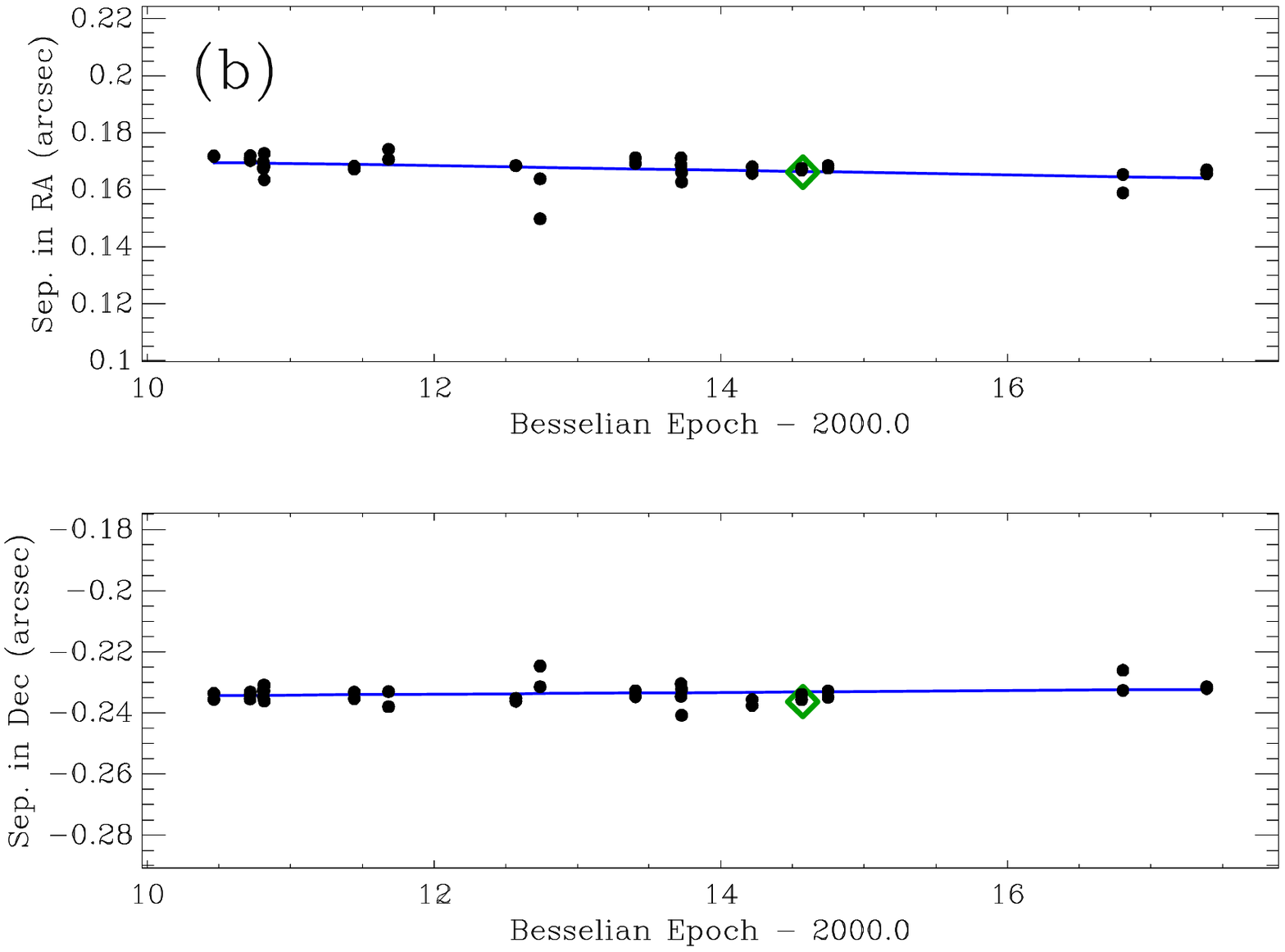}

\vspace{0.5cm}
\plottwo{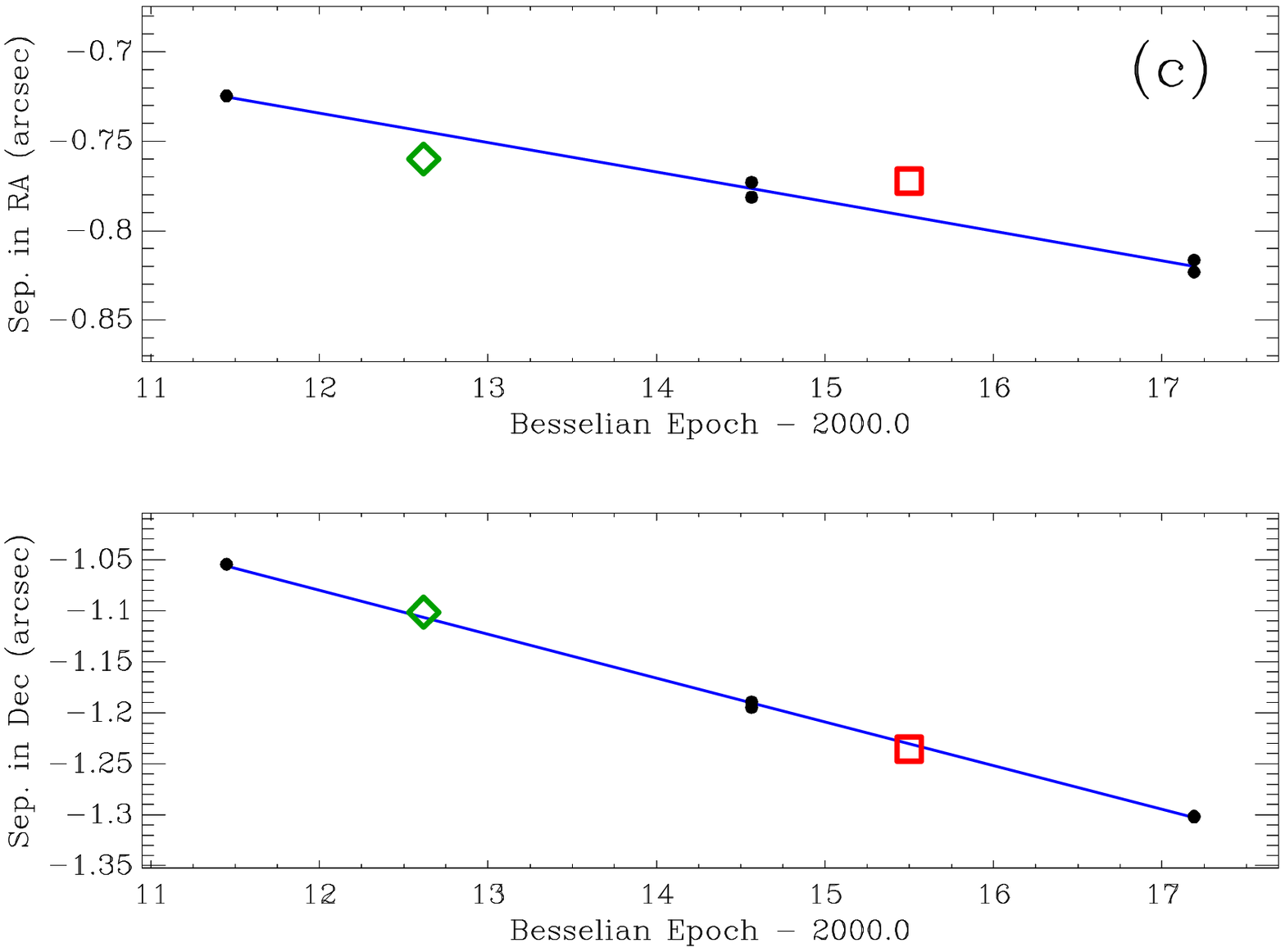}{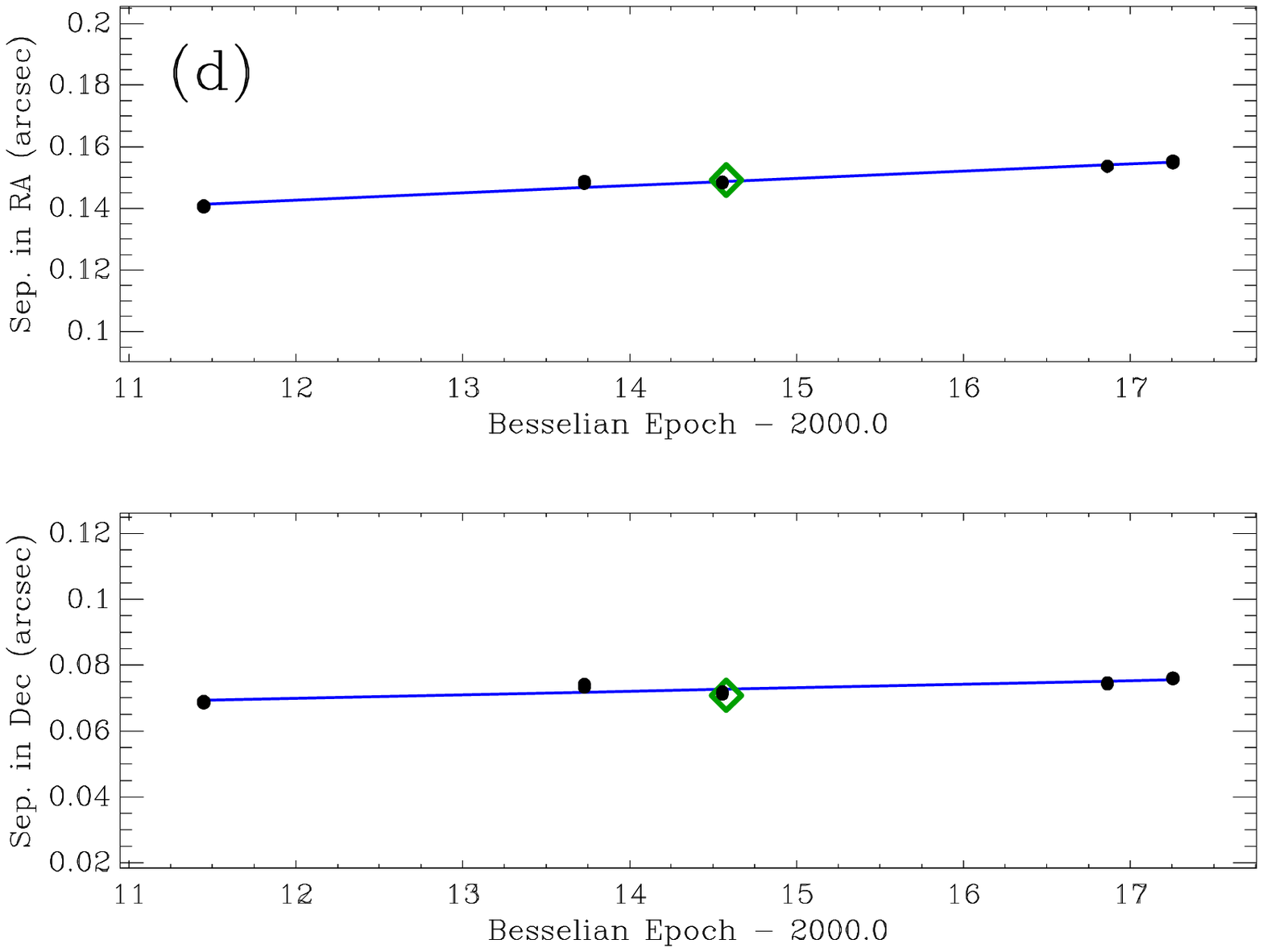}
\figcaption{Plots of the projected separation as a function of time for four representative stars in Table 2. 
Filled circles represent data points from Table 1, the red open squares mark the {\it Gaia} DR2 measurement (at J2015.5) if available, and the green diamonds are measures from \citet{kra16}.
The linear fit for each system, obtained as described in the text, is shown in blue, yielding the 
relative proper motion values appearing in Table 2. (The fits do not include the {\it Gaia} or 
Kraus et al.\ measures.) (a) KOI 13 = Kepler-13. (b) KOI 98 = Kepler-14. These two 
systems illustrate our results on extremely well-observed pairs. (c) KOI 118 = Kepler 467. (d) KOI 270 = Kepler 
449. These latter two examples show systems with the minimum or near-minimum number of observations to
be included in Table 2.}
\end{figure}

\begin{figure}[!htbp]
\figurenum{4}
%\plotone{f4.eps}
\hspace{4.5cm}
\includegraphics[scale=0.50]{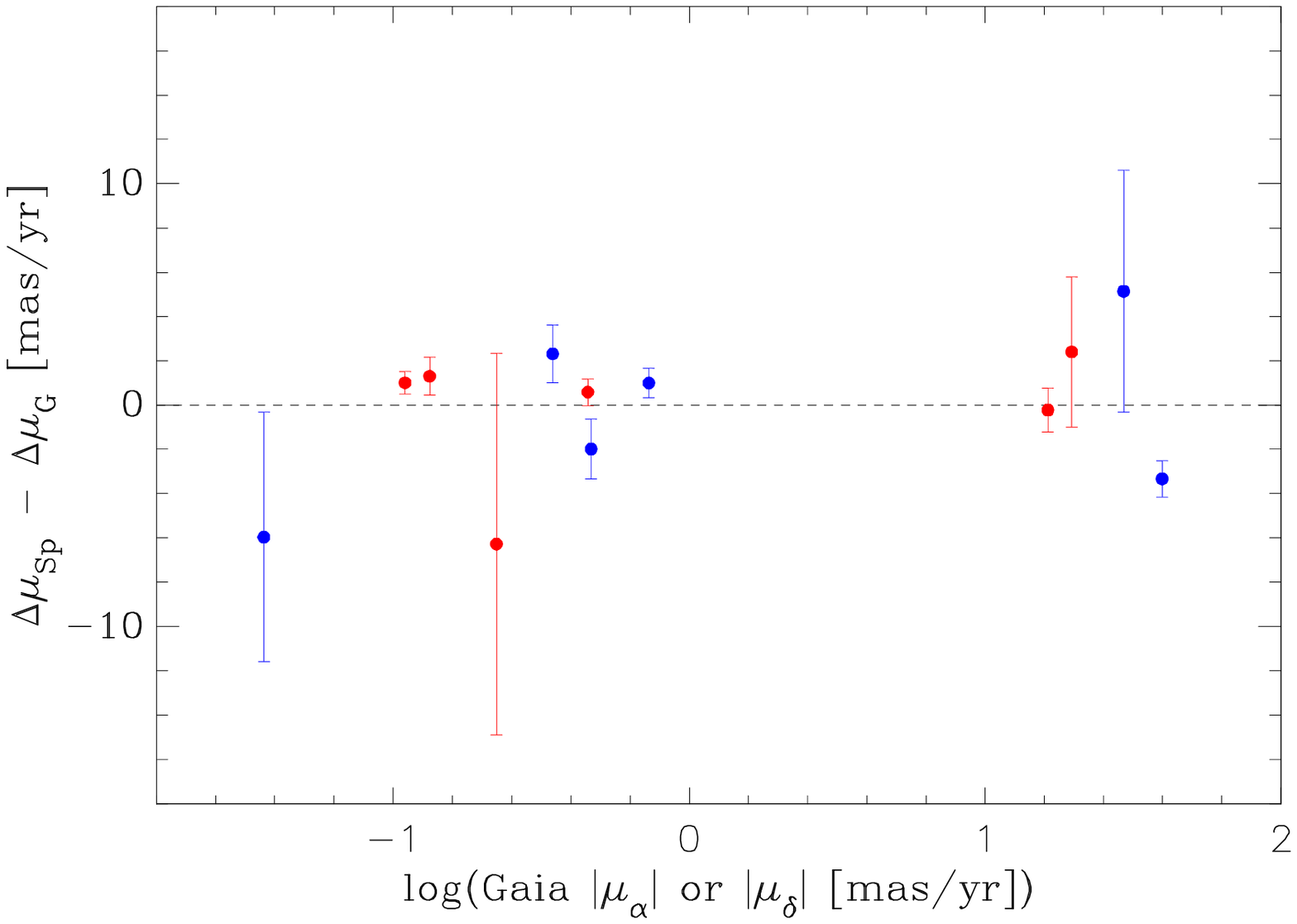}

%\figcaption{
\caption{
A comparison of the relative proper motions derived from the speckle data set versus those
in the {\it Gaia} DR2 for the six systems in Table 2 for which {\it Gaia} values exist. The ordinate is the 
(speckle minus {\it Gaia}) difference in $\Delta\mu_{\alpha}$, shown in red, and in $\Delta\mu_{\delta}$, shown in blue.}
\end{figure}

Given that most objects in Table 1 have observations at three or more epochs, we can use the positional
information over time to the determine relative proper motion of the secondary with respect to the primary. 
For this purpose, we compute
the separation projected onto the right ascension and declination system at each epoch. We perform
a linear least-squares fit to the separation in each coordinate as a function of time, where all data 
points receive  equal weight. From this, we can derive the slope of the line and its uncertainty, which 
represents the relative proper motion, $\Delta \mu_{\alpha}$ for right ascension, and $\Delta \mu_{\delta}$ 
for declination.  Table 2 gives results for  
$\Delta \mu_{\alpha}$ and $\Delta \mu_{\delta}$ for all objects in Table 1 that were observed at three or more
different epochs, and Figure 3 shows the fits obtained in four representative cases. 
We also plot there the positions from the {\it Gaia} DR2 \citep{gai18} in two cases,
and from \citet{kra16} in all four cases. We have compared the linear  
fits with and without the Kraus et al.\ data, for example, but see little difference in  
the uncertainties of the derived proper motions. To guard against introducing any systematic offsets
in position from different studies, we include only our own measures in deriving the proper motions.
Also in Table 2, we show the proper motion of the primary star (if resolved) or the system (if
unresolved) as it appears in the literature, generally either from the  {\it Gaia} DR2, Tycho-2 
\citep{hog00}, or UCAC4 catalogs \citep{zac13}. 
We also include the best known parallax, $\pi$, for the object from the literature. 

With regard to 
the DR2 proper motions and parallaxes, the sample represents a range of uncertainty values in these
quantities, with the least precise values comparable to the ground-based sources we have used when DR2
values are not available. We examined the Renormalized Unit Weight Error (RUWE) values for those objects in our sample, and as expected,
a majority of our stars have a high value of this quantity, which can indicate unresolved binaries
and other problems with the astrometric solution. Nonetheless, the RUWE values scale roughly
linearly with uncertainties in proper motion and parallax for our stars, and so we find no evidence
to suggest that the DR2 values are not reliable to the level that their uncertainties indicate.

There are six cases from Table 2 where both components are measured separately in DR2, all of which have
separations larger than one arc second (due to the resolution limitations of DR2). In Figure 4, we
compare our relative proper motion for the secondary versus what can be derived from the proper motion
of both components from DR2. We find no systematic offset in either $\Delta \mu_{\alpha}$ or $\Delta \mu_{\delta}$,
giving confidence that the methodology used on the speckle data is sound.

\section{Analysis of the Motions}

We investigate the motions derived in the previous section in two ways. First, one can compare the relative
motion of the components in our sample with what would be expected for line-of-sight (LOS) companions. In 
this case if the motion we observe is small compared to the mean motion of optical doubles, then the pair 
may be considered to be co-moving, or in other words, a common proper motion (CPM) pair, and thus likely 
to be physically associated. On the other hand, we can also study the motions we observe in comparison to 
the expected motions of orbiting companions with similar distances and masses. If the relative proper motion 
is consistent with the typical motion one would observe from a bound companion, then that can be used as 
further evidence of the nature of the system. While neither approach is definitive in terms of characterizing the 
companions as bound or unbound, both together can give a strong indication of the physical proximity of
the two stars.

\subsection{Common Proper Motion}

The stars in our sample cover a substantial distance range, from roughly 100 to 1800 pc, and the average proper motion
will decrease as the distance grows. Therefore, we form the ratio of the 
magnitude of the relative proper motion vector we derive to the magnitude of the system proper motion vector. 
We will refer to this ratio as $R_{1}$, so that
\beq
R_{1} = \frac{|\Delta \vec{\mu}|}{|\vec{\mu}_{1}|},
\eeq

\noindent
where $|\Delta \vec{\mu}|$ is the magnitude of the relative proper motion vector and $|\vec{\mu}_{1}|$ is the 
magnitude of the system proper motion vector, or that of the primary star if available.
If the pair exhibits CPM, then $R_{1}$ should be near zero, while if the companion is a distant 
background star, then the ratio will be near unity, since
the background star's proper motion in general will be much smaller than that of the primary star. Alternatively,
if the companion is a dim foreground star, the ratio is more likely to exceed unity, since in that case the
relative proper motion will be dominated by the motion of the secondary star. A pair of unknown disposition
can be judged as likely to be a CPM pair if its ratio is small, and more likely to be an optical double if the ratio is near
one or above.  

To understand the expected properties of this ratio for our sample, we chose two 
samples of 2000 stars 
from the {\it Gaia} DR2 catalogue near the center of the {\it Kepler} field. The first sample was selected to
have $G$ magnitudes between 10 and 14 and parallaxes greater than 0.555 mas, the smallest
parallax in our sample of KOIs. The second sample
had $G$ magnitudes between 10 and 18 with no parallax restriction. 
A small number of stars were common to both samples, and these were removed from consideration, leaving
two independent samples of 1944 stars each.
We then form a hypothetical optical
double by randomly pairing a star from the first sample to act as a primary star and a star from the second
sample to act as a secondary. The DR2 data was used to compute the relative proper motion between
the stars, and we form the ratio $|\Delta \vec{\mu}|/|\vec{\mu}_{1}|$ using the primary's proper motion. 
We select
from the sample only systems with magnitude differences between 0 and 4.6 (roughly matching our expected
detection limit) for inclusion in the final sample. This left a total of 1366 optical pairs, from which we can study
the statistics. The ratio $R_{1}$ is plotted as a function of distance for the final
sample in Figure 5(a). As expected, the sample has a median value near 1, indicating that in most cases
the companion is a background star that moves little in comparison to the closer primary star; hence, the
relative proper motion is nearly the same value as $|\vec{\mu}_{1}|$. 

Foreground companions (which by construction
are always dimmer than the primary) represent 11.7\% of
the cases in the final sample overall and have a higher median value, near 1.6. Indeed, if only stars above
the median line in the plot are considered, 
foreground companions then represent 15.5\% of the sample. 45.6\%
of companions above the 95\% line are foreground stars.

\begin{figure}[!t]
\figurenum{5}
\plottwo{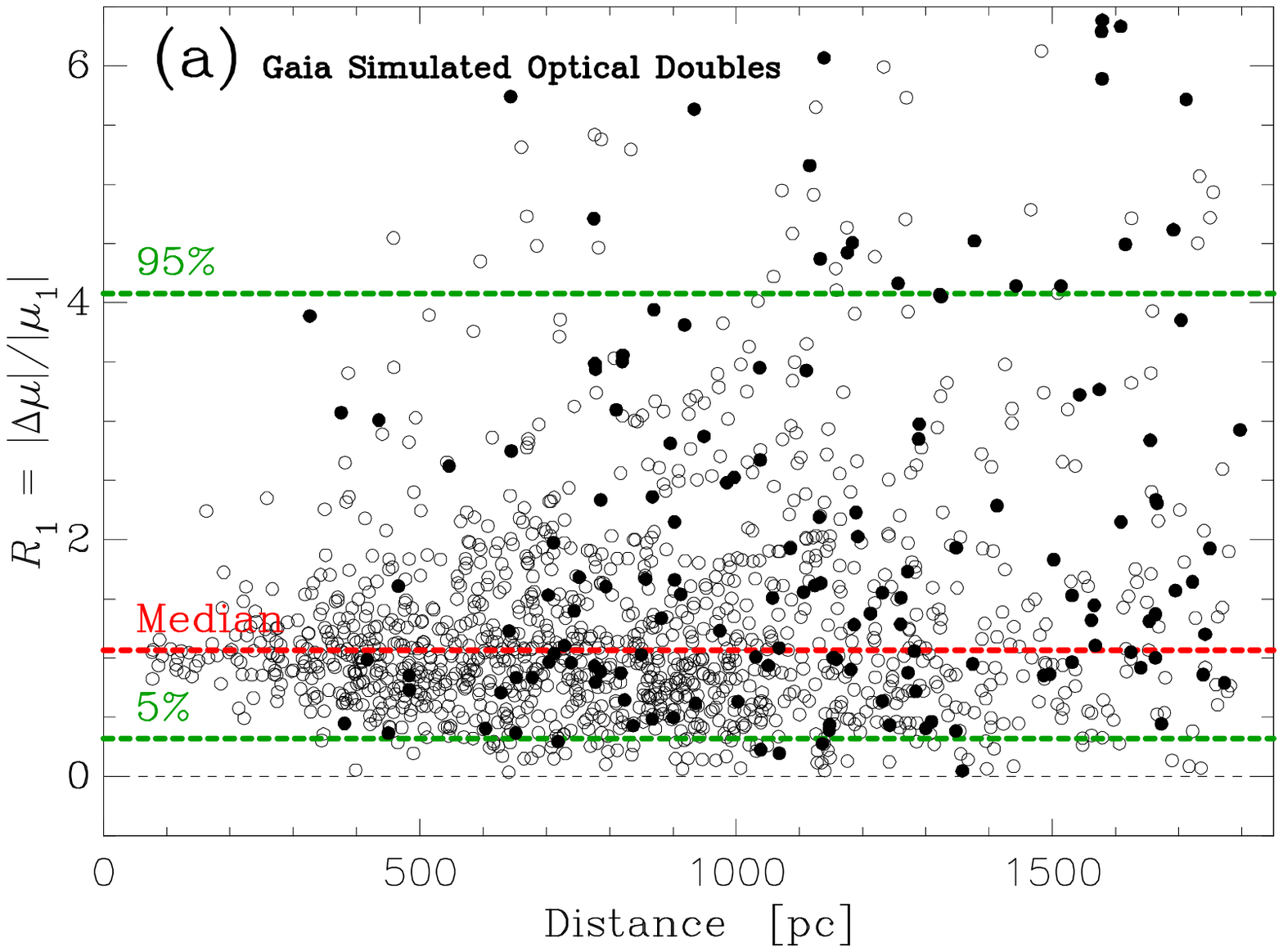}{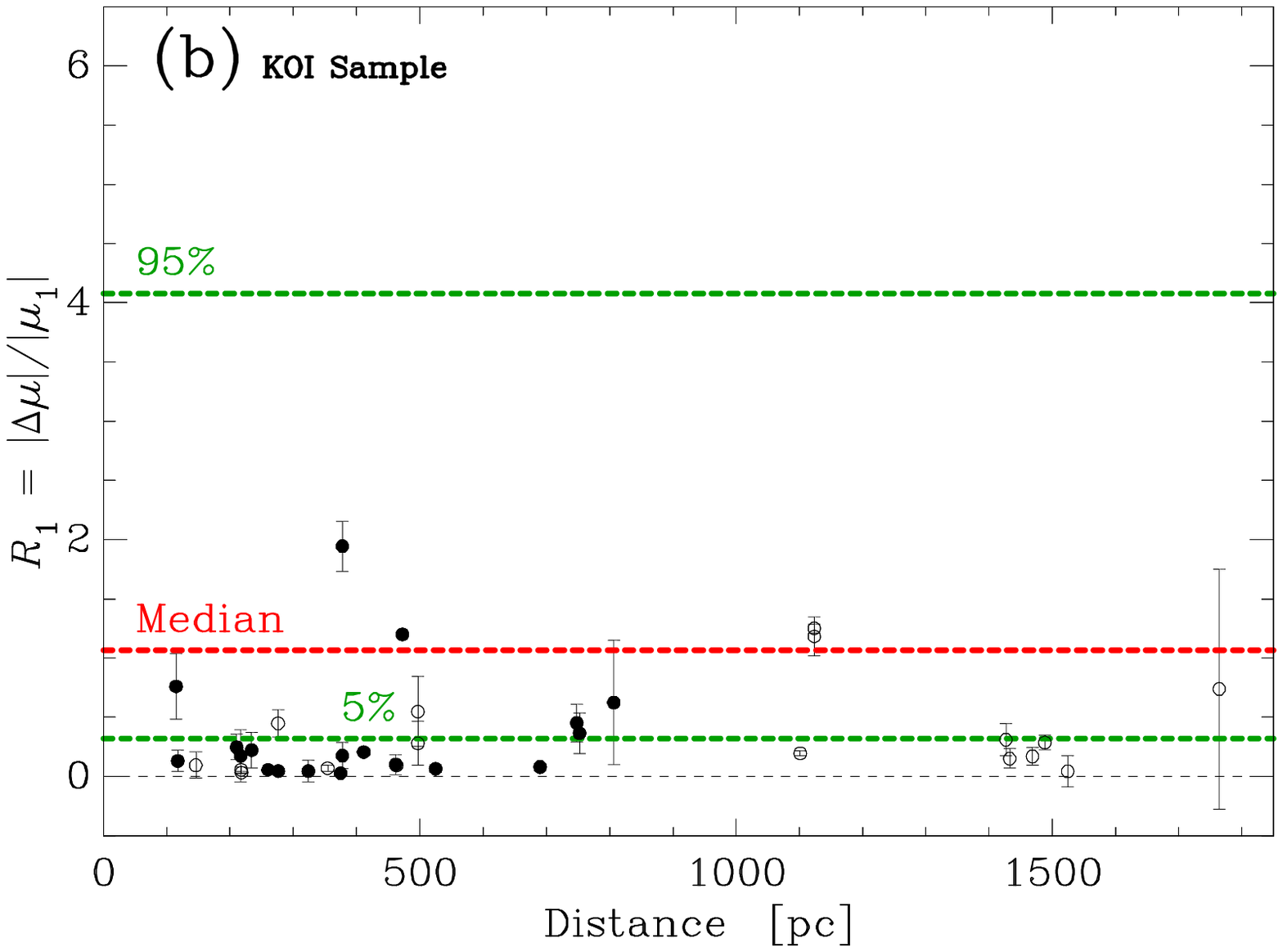}
\figcaption{The ratio $R_{1}$ (the magnitude of the difference in proper motion between primary and secondary
divided by the proper motion of the primary) as a function of distance. (a) In this panel, we show stars from the 
{\it Gaia} DR2 catalog located near the center of the {\it Kepler} field that have been selected and randomly paired with a second star
to approximately match the magnitude differences of the stars in Table 2. The primary stars have distances
in the range of 0 to 1800 pc and {\it Gaia} $G$ magnitudes between 10 and 14. This simulates a population
of line-of-sight companions as discussed in the text. Foreground companions are shown as filled circles,
and background companions as open circles. (b) Here, we plot the same
for the stars in Table 2, where filled circles indicate systems with parallax measures with uncertainties under
10\%. In both panels, the dashed lines indicate the percentage of stars in (a) that have a 
ratio below the line drawn, thus the lower green line is the 5\% contamination line of optical doubles.}
\end{figure}

The results in the plot also show that only
5\% of hypothetical optical doubles have $R_{1} < 0.32$ over the distance range of our sample.
In Figure 5(b), we show the same plot for the KOI sample. The majority of stars have ratios below 0.32, and
thus may be viewed as likely to be common proper motion pairs. We also find three of our systems (including
one triple, KOI 1792) above the median line; these are almost certainly not CPM pairs. KOI 1792 will
be discussed further in Section 7.

Although the majority of our systems appear likely to be CPM pairs, it is wise to remain cautious at this stage
concerning a determination. Figure 5(a) illustrates that, although
unlikely, chance similarities in proper motions can exist that will make a physically unbound system appear
to be a CPM pair, particularly for distances larger than 300 pc, which is a large majority of our sample.

\subsection{Orbital Motion}

For the stars in our sample, we do not expect large changes in the relative position of the components 
due to orbital motion over the time that we have observed them, given the
range of distances represented and the likelihood of very long periods, if bound. 
Nonetheless, as a simple example, a 2$M_{\odot}$ system with a 
semi-major axis of 
500 AU in a face-on circular orbit at a distance of 1 kpc will have a period of 7900 
years and an observed separation of 0.5 arcsec, and therefore move about 0.4 mas/year
on the sky, a number that is comparable to many of the uncertainties stated for the relative proper motions in 
Table 2. For that reason it can be expected that in at least some cases, the orbital motion will be 
above the measurement precision represented by our data set.

To investigate this, we define the quantity $R_{2}$ as follows:

\beq
R_{2} = \frac{|\Delta \vec{\mu}|}{|\Delta \vec{\mu}({\rm avg,orb})|},
\eeq

\noindent
where $|\Delta \vec{\mu}|$ is again the magnitude of the relative proper motion vector between the 
primary and secondary star, and the quantity $|\Delta \vec{\mu}(\rm avg,orb)|$ is an estimate of the 
expected average change in relative position per year assuming
that the system is gravitationally bound. As with the previously defined quantity $R_{1}$, creating a 
ratio has the virtue of removing the decrease in the observed angular motion with increasing 
distance. We can also 
anticipate that if the denominator is estimated well, then the average value of $R_{2}$ for bound 
companions will be near 1. If the system is a line-of-sight companion, then $R_{2}$ can scatter toward 
values that are significantly larger than 1 in general since in that case the proper motion of either star 
can be a few to 10's of mas per year (as shown in Table 2). This is much larger than the typical orbital motions, 
and the numerator
of $R_{2}$ will therefore be much larger than the denominator.

We estimate the quantity $|\Delta \vec{\mu}({\rm avg,orb})|$ by starting with the effective temperatures for the KOIs in 
Table 2 as they appear in the Exoplanet Follow-up Observing Program (ExoFOP) website for {\it Kepler}\footnote{\tt https://exofop.ipac.caltech.edu/cfop.php}. In
many cases there are estimates made by multiple observers, and in those cases the average values were calculated,
together with the standard errors. Individual effective temperature measures often have uncertainty estimates, and in
general we find that these are consistent with the values derived by computation of the standard error. On average,
the errors we obtain in the temperatures in this way are roughly 100K. 

Next, the temperatures are used to estimate the mass, spectral type, and absolute $V$ magnitude 
of the primary star. 
A standard reference, \citet{sch82}, is used for that purpose. We then take the average magnitude 
difference for each KOI in Table 2 in the 562-nm filter and we use that to estimate the absolute $V$
magnitude of the secondary star, on the assumption that it is at the same distance as the primary star 
(i.e.\ that the system is bound). \citet{sch82} is then used to arrive at the mass of the secondary star
and a total mass estimate for the system is 
computed. Two stars in our sample, KOI 258 and KOI 977, are listed as giants in 
SIMBAD\footnote{\tt https://simbad.u-strasbg.fr/simbad}
\hspace{0.05cm} \citep{wen00}; this was taken into
account in computing the masses, but otherwise primary stars are assumed to be dwarfs.

Finally, using the observed separation as a proxy for the semi-major axis of the orbit, we compute a period estimate $P$ according to Kepler's harmonic law:

\beq
P = \sqrt{\frac{\rho^{3}}{\pi^{3} M_{\rm tot}}},
\eeq

\noindent
where $\rho$ is the separation of the pair, $\pi$ is the parallax, and $M_{\rm tot}$ is the estimated
total mass of the system.
The estimate of $|\Delta \vec{\mu}({\rm avg,orb})|$ is then obtained by computing the total distance traveled in one orbit
on the plane of the sky, that is, computing the perimeter of the observed orbital ellipse, and dividing that 
by the period itself. This gives an average distance covered per year, or equivalently, an average value for 
the observed relative proper motion due to orbital motion.
The perimeter of the ellipse will vary with the various orbital elements, in particular the inclination
$i$, the eccentricity $e$, and the ascending node $\Omega$, which of course are not known. However, the perimeter should scale with the semi-major axis and be dependent on orbital geometry; for 
an edge-on circular orbit of semi-major axis $a$, the total distance on the sky  that is traveled in one orbit is $4a$. On the other
hand, for a face-on circular orbit, the result is $2 \pi a$. Thus, we can infer that for any other orbit, the total
length of the orbit in the plane of the sky is a geometrical factor times the semi-major axis, and that the factor is of 
order 5. Using the initial mass function found in \citet{kro01}, we constructed a simulation of 5000 primary stars and then chose binary companions and orbital parameters using \citet{2010ApJS...190...1R}. 
The length of the perimeter of each orbit in the plane of the sky was then calculated,
and compared to the semi-major axis. This exercise showed that, on average, the value of the geometrical factor
is approximately 4.6. Therefore we propose to calculate $R_{2}$ from observational data using the following method:

\beq
R_{2} =  \frac{|\Delta \vec{\mu}|}{|\Delta \vec{\mu}({\rm avg,orb})|} = 
\frac{|\Delta \vec{\mu}| \cdot P}{4.6 \cdot \rho} =
\frac{|\Delta \vec{\mu}|}{4.6} \sqrt{\frac{\rho}{\pi^{3} M_{\rm tot}}}
\eeq

We used a sample of simulated binaries to inform us how our proposed quantity $R_{2}$ behaves 
for the KOIs in Table 2. Again employing our simulation of 5000 orbits, and we
assign distances with a Gaussian distribution
with the same mean and standard deviation as the sample of stars in Table 2. We check to make sure that the 
apparent magnitude, separation, and magnitude difference of each system falls in the range that we could observe 
with DSSI. Finally, we compute a value of $|\Delta \vec{\mu}|$ at a random point in the orbit, and form $R_{2}$ 
as shown in Equation 5.
The results are shown in Figure 6(a) and they indicate that these simulated binaries have $R_{2}$ values that 
cluster
around 1 regardless of distance, and that only 5\%  of the orbits generate an $R_{2}$ value greater than 
1.87. 
Computing the value of $R_{2}$ for our data, we obtain the plot in Figure 6(b). In this case we see that the majority of
our stars have motions that are consistent with orbital motion, but that some do have values much higher than 
1.87, which indicates that the relative motion cannot be explained by mutual gravitational attraction, 
at least given the mass 
estimates used. We investigated how sensitive the ratio is to the mass sum of the simulated binaries 
and found that if masses twice as large were
used in generating the orbits used in the simulation, the 95\% line drawn
in Figure 6 would increase to 1.80 (the log of which is 0.26). 
Given the range of $R_{2}$ values of the KOI stars we have studied, 
this is not a large change. Thus we conclude that our results are not strongly dependent on the mass 
estimates we have made.

\citet{ker19} use a similar methodology to determine the physical association of galactic Cepheids and RR Lyrae stars in {\it Gaia} DR2; however, in their case they use the total mass and separation to compute
an estimate of the escape velocity of the system. This is then compared with the relative proper motion.
We note that the denominator in Equation 5 has the same dependence on total mass and separation as the standard escape velocity formula, namely 

\beq
v_{\rm esc} = \sqrt{\frac{2G M_{\rm tot}}{r}} \sim \sqrt{\frac{M_{\rm tot}}{r}},
\eeq

\noindent
and therefore our method is based on essentially the same physics, and at most differs in the exact value of
$R_{2}$ that is used to differentiate between bound and unbound companions; however, our method has more detail in the sense that known statistics of the orbital parameters of field binaries are incorporated into the model.

\begin{figure}[!t]
\figurenum{6}
\plottwo{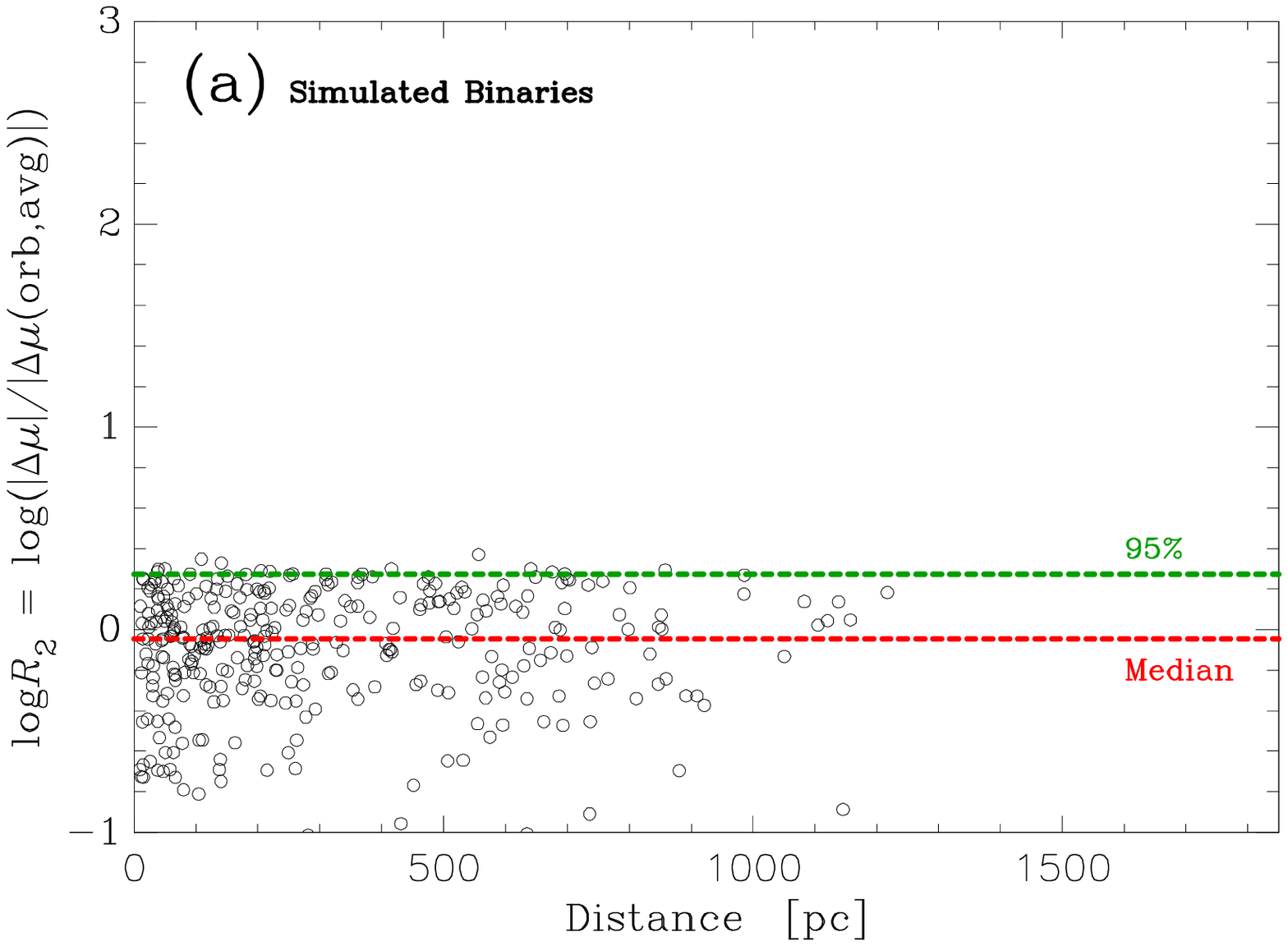}{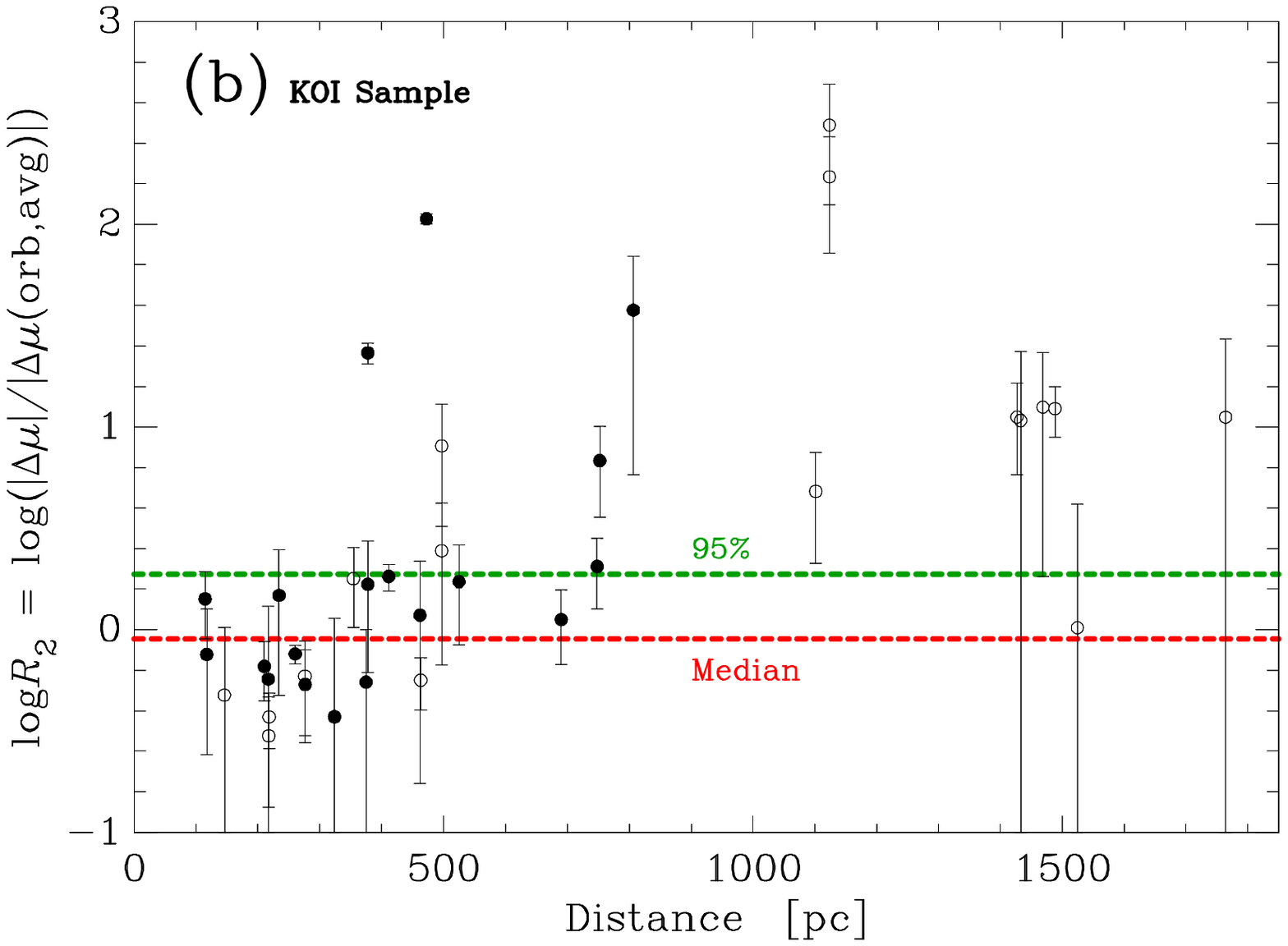}
\figcaption{The log of the ratio $R_{2}$ (the magnitude of the difference in proper motion between primary and secondary
divided by the derived estimate of the average relative proper motion based on the assumption of orbital motion), plotted
as a function of distance. 
(a) In this panel, we show the simulated distribution created as described in the text for bound companions. 
(b) Here, we plot the same for the stars in Table 2. In both panels, the dashed lines indicate the percentage 
of stars in (a) that have a ratio below the line drawn.}
\end{figure}

\section{Properties of the Sample}

Table 3 contains the final information for our 34 well-measured systems, showing columns for (1) KOI number; 
(2) Kepler number if any; (3) the number of confirmed exoplanets in the system, (4) the 
distance to the system; (5) the estimate for the primary mass; (6) the estimate 
for the secondary mass; (7 and 8) the values for $R_{1}$ and $R_{2}$ obtained from 
the analysis in the previous 
section; and (9) final comments regarding the disposition of each system. 
If a {\it Gaia} DR2 parallax is available, the distance value in Column 5 is that in \citet{bai18} 
rather than simply inverting the {\it Gaia} DR2 parallaxes. The differences in all cases except 
KOI 1890 are small, generally within a few percent, with no systematic trend for our small sample.

From this, we can conclude that the 
majority of these systems are highly likely to be gravitationally bound, and that a few are obvious 
LOS companions. The final disposition for each system, shown in the rightmost column, is determined
from $R_{1}$ and $R_{2}$ as follows:

\vspace{0.2cm}
\noindent
$\bullet$ A system is judged to be a CPM pair if $R_{1} + \delta R_{1} < 0.32$ and 
$R_{2} - \delta R_{2} <  1.87$. This occurs when (1) the full uncertainty interval for $R_{1}$
falls below the 5\% line in Figure 5(b) so that we can confidently claim that the pair is co-moving
and the (2) uncertainty interval for $R_{2}$ overlaps with the 95\% probability line in Figure 6(b),
indicating that the motion is consistent with at least some fraction of orbital possibilities for the system.

\begin{deluxetable}{rrcrrrrrl}
\tabletypesize{\small}
\tablewidth{0pt}
\tablenum{3}
\tablecaption{Final Properties of 37 KOI Double Star Components}
\tablehead{
\colhead{KOI} &
\colhead{Kepl. No.} &
\colhead{No. of} &
\colhead{Distance} & 
\colhead{$M_{1}$\tablenotemark{b}} &
\colhead{$M_{2}$\tablenotemark{b,c}} & 
\colhead{$R_{1}$} &
\colhead{$R_{2}$} &
\colhead{Comments} \\
\colhead{No.} &
\colhead{or Disp.\tablenotemark{a}} & 
\colhead{Planets} &
\colhead{(pc)} &
\colhead{($M_{\odot}$)} &
\colhead{($M_{\odot}$)} &
&&
}
\startdata
1 & 1 & 1 & $ 215.3^{+1.1}_{-1.0}$ & 0.97 & 0.51 & $0.174 \pm 0.223$ & $0.57 \pm 0.73$ & CPM? \\
13 & 13 & 1 &  $519.1^{+30.7}_{-27.5}$ & 1.60 & 1.48 & $0.066 \pm 0.033$ & $1.73 \pm 0.89$ & CPM \\
98 & 14 & 1 & $690.1 \pm 55.7$ & 1.40 & 1.12 & $0.082 \pm 0.032$ & $1.12 \pm 0.45$ & CPM \\
118 & 467 & 1 &  $466.3^{+5.1}_{-5.0}$ & 0.89 & (0.46) & $1.200 \pm 0.018$ & $106.16 \pm 5.99$ & LOS \\
120 & PC & ... & $789.4^{+28.5}_{-26.6}$ &  1.12 & (1.00) & $0.625 \pm 0.527$ & $37.60 \pm 31.78$ & Uncertain\\
%177 & (PC) & $0.716 \pm 0.118$ & $0.238 \pm 0.161$ & $5.5 \pm 3.5$ & $6.1 \pm 1.4$ & 
%258AB & (FP) & $1.089 \pm1.118$ & $1.012 \pm 0.675$ & $4.183 \pm 0.052$ & $-7.309 \pm 0.055$ &
%258AC & (FP) & $-15.745 \pm1.597$ & $-4.521 \pm 3.283$ & $4.183 \pm 0.052$ & $-7.309 \pm 0.055$ &
\enddata
\tablenotetext{a}{If no Kepler number is given, the disposition as either a planetary candidate (PC)
or false positive (FP) is given, based on information available on the Kepler CFOP website.}
\tablenotetext{b}{The uncertainty in all values in this column is assumed to be 0.1$M_{\odot}$.}
\tablenotetext{c}{Calculated under the assumption that the companion is bound. If the system
is not judged to be a CPM pair then the value is shown in parentheses.}
\tablecomments{Table 3 is published in its entirety in the machine-readable format.
      A portion is shown here for guidance regarding its form and content.}
\end{deluxetable}

\vspace{0.2cm}
\noindent
$\bullet$ A system is judged to be a probable CPM pair (labeled ``CPM?'' in Table 3) 
if $R_{1} < 0.32$ and $R_{2} - \delta R_{2} < 1.87$, and it is not already in the 
previous category. This relaxes the restriction on $R_{1}$ so that the measured value itself 
has a less than 5\% probability of being due to random, unrelated motion, but the uncertainty
interval includes higher probabilities.

\vspace{0.2cm}
\noindent
$\bullet$ A system is considered to be a LOS companion if $R_{1} - \delta R_{1} > 0.32$ and
$R_{2} - \delta R_{2} > 1.87$. 
In this case, the full uncertainty interval for $R_{1}$ is consistent with random
motions between the two stars, and the full interval of $R_{2}$ is too high to be consistent
with orbital motion.

\vspace{0.2cm}
\noindent
$\bullet$ A system is labeled as ``uncertain" if it does not fit into one of the above three categories. One
other system is also labeled as uncertain, KOI 959; this object has no parallax value
in the literature, 
and therefore we cannot form $R_{2}$ to complete the analysis here. However, judging from the 
value of $R_{1}$ and its uncertainty, the system is likely to be a CPM pair. The proper motion is
large, and the spectral type listed in SIMBAD is M+M. The system may therefore be quite nearby.
%For KOI 2837, no system proper motion
%appears in the literature, and so we cannot compute an $R_{1}$ value for our analysis. However,
%the motion we observe from the speckle data indicates a consistency with orbital motion.
%In future {\it Gaia} data releases, it will be important to see if the proper motion has been 
%included.

\vspace{0.2cm}
The placement of our systems in a $\log R_{2}$ versus $\log R_{1}$ diagram
is shown in Figure 7(a). Of the 37 components, 17 are CPM, 4 are probable-CPM, 4 are LOS, and the 
remaining 12 are
of uncertain disposition. Of those systems with confirmed exoplanets, 10 are CPM, 2 are probable-CPM, 2 are 
LOS, and 4 are uncertain.  Neglecting the uncertain systems,  these latter statistics show that
the fraction of CPM systems for the exoplanet host systems in the sample is 12 of 14, or $85.7 \pm 9.4$\%, 
which is consistent with the percentage predicted in \citet{2014ApJ...295...60H}, where 96\% of systems 
discovered at WIYN and 84\% of those discovered at Gemini were predicted to be gravitationally bound. 
Looking at the magnitude differences and separations of the three LOS companions, all of these fit into
the region where LOS companions are expected to be found in Figures 8 and 9 of \citet{2014ApJ...295...60H}. 
This is shown in Figure 7(b). Five systems with more than one confirmed exoplanet are in our 
sample: KOI 270, KOI 279,  KOI 284, KOI 307, and KOI 1792. Of these, three are CPM systems, 
1 is LOS, and 1 is uncertain.

These results may be compared with other studies  of KOI double stars already in the literature
that make a determination of whether the pair was bound or unbound using other methodologies. 
\citet{atk17} and \citet{zie18} both make a judgement using a photometrically-determined distance for both 
components; if the distance derived for the two stars is consistent within the uncertainty, then they infer 
that it is highly likely that the pair is physically associated. They both compute a value to assess the
probability of being bound based on the differences in the two distances in terms of their estimated 
uncertainties, the quantity they call ``$\sigma_{\rm unbound}$,'' with larger values implying a smaller 
chance that the system is physically associated. In the case of \citet{atk17}, although seven systems are 
common to both their study and ours, we judge three of these to be uncertain, and so there are only 
four cases where a clear comparison can be made, namely KOI 984,
1613, 2059, and 5578. Both studies find that the first three are all CPM or CPM? systems, but
we find that KOI 5578 is CPM while Atkinson et al. find a $\sigma_{\rm unbound}$ value above 3,
higher than what would be expected for a bound system. For \citet{zie18},
six systems are common to both studies, one of which we judge to be uncertain. We find clear 
agreement in three cases (KOI 1, 1792, and 2059), and apparent
discrepancies in two others (KOI 13 and 984). Again, we find these are likely bound while 
Ziegler et al.\ find $\sigma_{\rm unbound}$ values above 3.

The greatest overlap between our results and previous work is found in \citet{2017AJ...153...117H}. 
Those authors also 
attempted to determine whether the companions to KOI stars
are bound through photometric means, but in their case via H-R diagram placement of the components. If the two stars were consistent with a common isochrone, the pair was judged to be bound. 
As with \citet{atk17} and \citet{zie18}, a direct comparison cannot be made in every case 
as either Hirsch et al.'s study or ours obtains an uncertain result,
but for the 19 companions where both studies make a determination, there is agreement in 17 cases 
(15 CPM or CPM? that Hirsch et al.\ judge as bound, and 2 LOS that Hirsch et al.\ judge as unbound).
In two cases, KOI 270 = Kepler-449 and KOI 984, Hirsch et al.\ found
that the system is unbound, yet our measured motions to date indicate a co-moving pair; 
in the case of KOI 984, we judge it to be a probable-CPM pair. This system has a large separation,
leading to potentially over-estimated magnitude differences in speckle observations
due to speckle decorrelation (as noted in 
the observations in Table 1), and so this might have affected the Hirsch et al.\ photometric result. 
For Kepler-449, this is not a concern as this object has a small separation, so it is possible in this case 
that there is simply a chance alignment in velocities to make the pair appear co-moving. However,
looking at the additional differential photometry available in Table 1 and combining that with what
Hirsch et al.\ had available earlier, we note that the preponderance of the evidence
points to a pair with modest magnitude difference at all observed wavelengths, which suggests that 
it would be possible to reconcile the photometry with a common isochrone for both stars, assuming
both stars are dwarfs. 
Four exoplanet host systems from our sample of particular note, 
including Kepler-449, are discussed below. 

\begin{figure}[!t]
\figurenum{7}
\plottwo{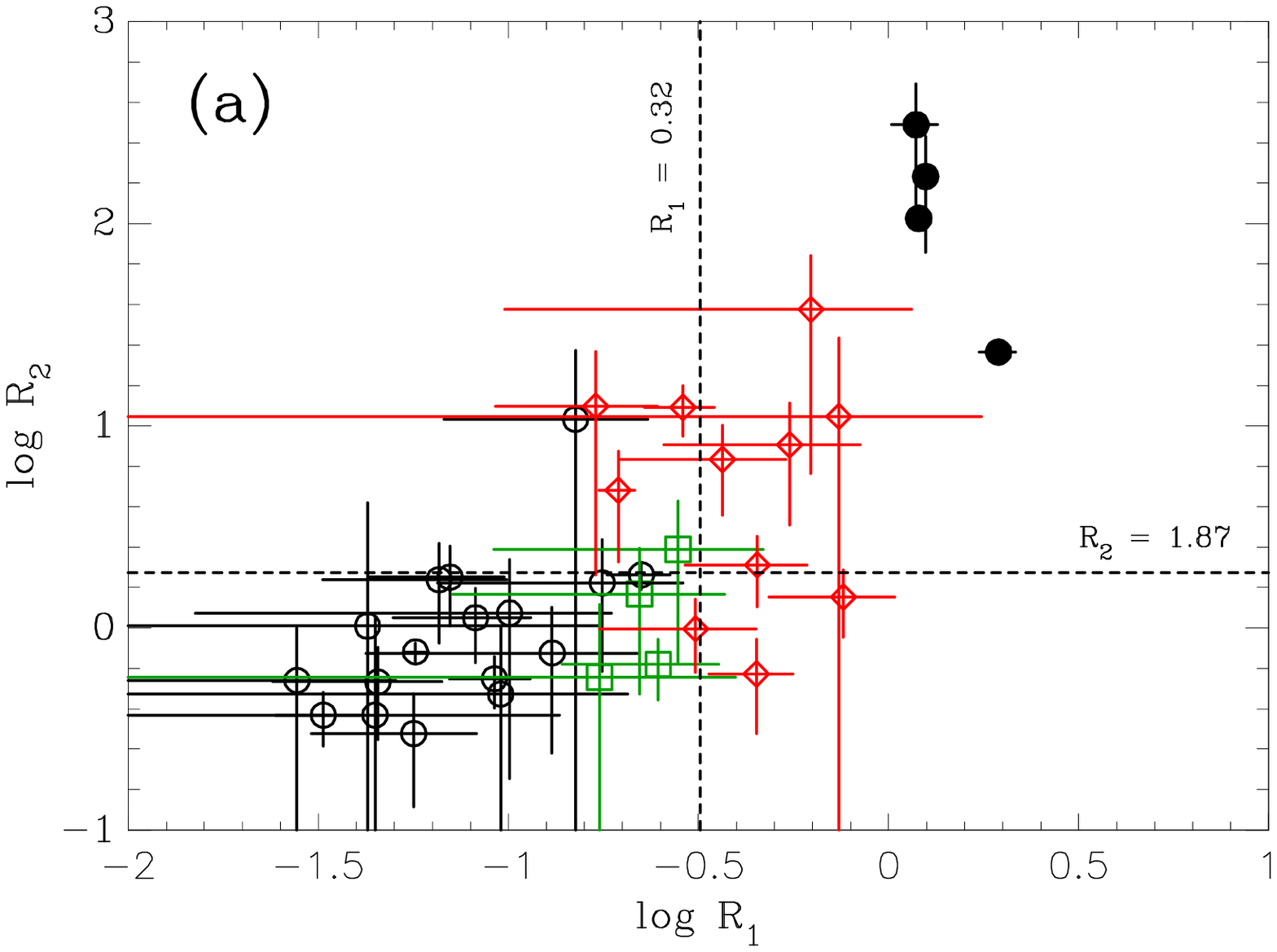}{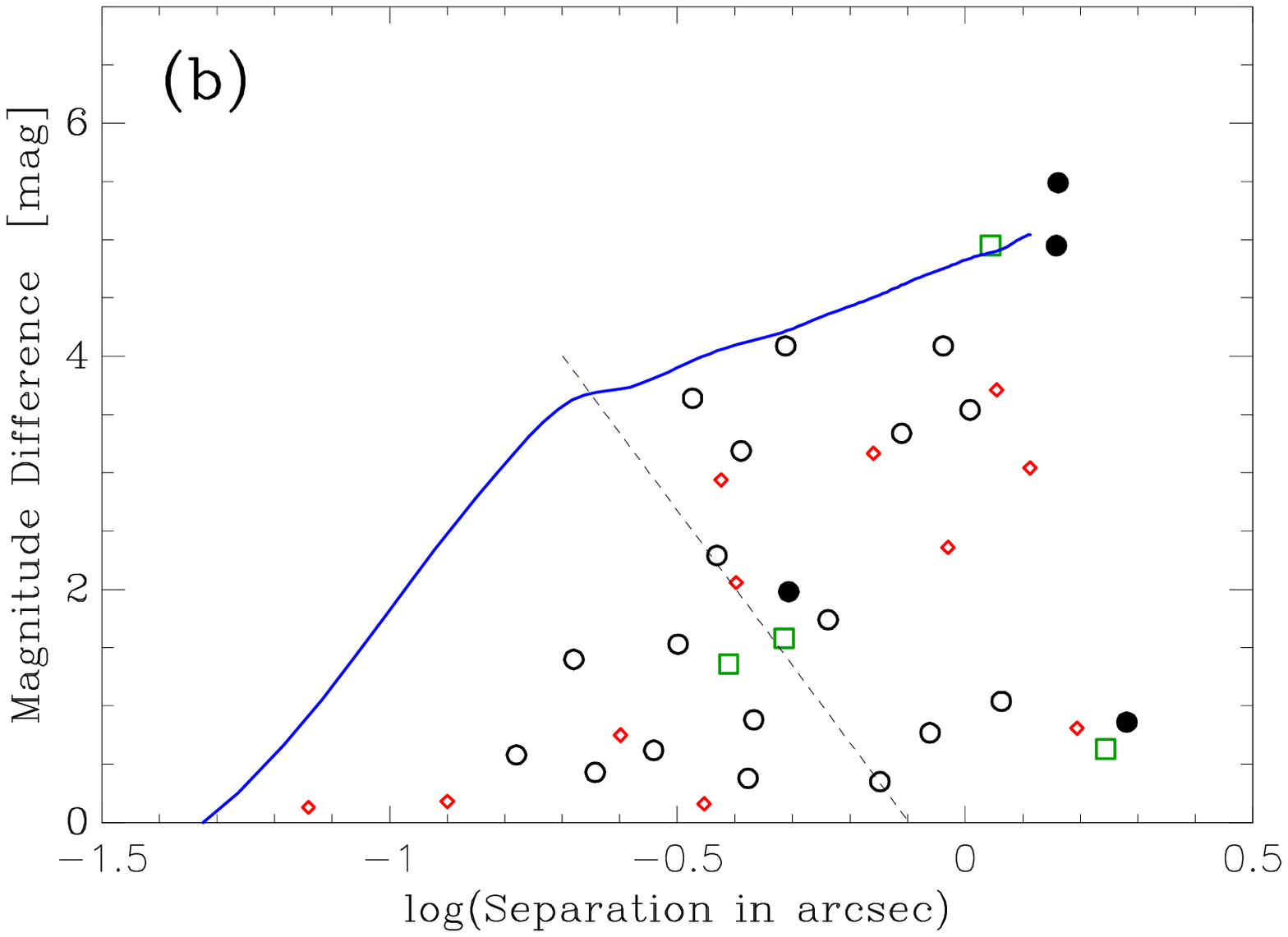}
\figcaption{
(a) A plot of $\log R_{2}$ as a function of $\log R_{1}$ for the 35 components that have both
values listed in Table 3. 
(b) A plot of magnitude difference as a function of separation for the components in Table 3.
In both plots, open circles are CPM pairs, open green squares are the probable-CPM pairs 
(listed at ``CPM?'' in Table
3), filled circles are LOS companions, and small red diamonds are the systems of uncertain disposition.
In panel (a), the values of $R_{1}$ and $R_{2}$ used to make the final determination for each system
are drawn as dashed lines; in general, we expect LOS companions to be in the upper right 
while CPM pairs are in the lower left.
In panel (b), the same detection limit curve as in Figure 1(b) is drawn, and the dashed line indicates the approximate demarcation between the region where LOS companions are not found (below the line), and the
region where there is a mix of bound and unbound companions (above the line) in the simulations detailed in 
\citet{2014ApJ...295...60H}.}
\end{figure}

%\hspace{4.5cm}
%\includegraphics[scale=0.50]{f7.eps}
%\caption{

\vspace{0.2cm}
\noindent
{\bf Kepler-13 (KOI 13).} The photometric analysis of \citet{2017AJ...153...117H} showed that the secondary star in this system 
(which has one confirmed exoplanet) is likely to be a bound companion. The preliminary 
astrometric analysis of \citet{hes18} reached the same conclusion, and with additional data, we also 
find that the relative proper motions confirm this determination, though as just mentioned, \citet{zie18}
reach a different conclusion. This system was also studied in detail by \citet{how19}, so assuming
its determination as CPM pair is correct, it is one of the few KOI binary systems for which it is 
known observationally that the exoplanet orbits the primary star and not the secondary. 

%\noindent
%{\bf KOI 258.} Although this system has been determined to be a false positive as an exoplanet host,
%this optical triple star consists of an F6III primary and two faint companions, one at
%a separation of approximately 1 arc second, and the other at nearly 1.5 arc seconds. The evidence
%from this analysis suggests that the closer companion is a CPM component, but the wider one is not.
%Thus it is most likely a binary star with projected separation of $\sim$380 AU that also has a somewhat
%wider LOS companion.

\vspace{0.2cm}
\noindent
{\bf Kepler-449 (KOI 270).} This solar-type star hosts two exoplanets 
with orbital semi-major axes of 0.1 and 0.2 AU respectively, 
both of which have current mass estimates below $2M_{\oplus}$. The companion star has a projected 
separation of 43 AU. If the stellar companion is orbiting in the same plane as the planets (i.e.\ the
inclination of the stellar orbit is near 90$^{\circ}$), then the minimum period for the stellar orbit (obtained
by assuming we are observing the maximum separation at the present time) would be
approximately 230 years, given the total mass estimate in Table 3. The motion of this object
is shown in Figure 3(d) and further discussed below; 
despite its relatively small separation and proximity to the Solar system compared with other stars in the
sample, we do not see obvious evidence for acceleration in our
observations at this point.

\vspace{0.2cm}
\noindent
{\bf Kepler-132 (KOI 284).} Similar to KOI 270, this star has effective temperature near that of the Sun, and is 
a multi-planet system, hosting four known exoplanets. Three have 
orbital semi-major axes of less than 0.2 AU, but the fourth orbits in a nearly circular orbit of
0.44 AU and period of 110 days. All have current mass estimates below $2M_{\oplus}$.
The stellar companion has a large projected separation compared
with these planetary orbits, 240 AU, but is less than one magnitude fainter than the primary star. 
Making the same calculation as above for the orbital period, we find that, if the stellar component
is co-planar, it would have an orbital period greater than 3000 years.

\vspace{0.2cm}
\noindent
{\bf Kepler-953 (KOI 1792)} This star hosts three confirmed 
exoplanets, and has $R_{1} > 5$
and $R_{2} > 100$ for both the B and C components. As can be seen in Table 2, both the B and C
components have the same relative proper motion to the primary star, thus these fainter two stars
are traveling together and form a CPM double. The large value of both $R_{1}$ and $R_{2}$ points 
to a slightly elevated probability of the fainter pair being interposed along our line of sight in the direction
of the exoplanet system, but on the other hand, the C component is resolved from AB in DR2, 
and it would appear to have a smaller parallax than AB, indicating that it would be a background pair.
Currently, the parallax uncertainty is too large to make a definitive statement from DR2 alone, however.

\begin{figure}[!t]
\figurenum{8}
\plottwo{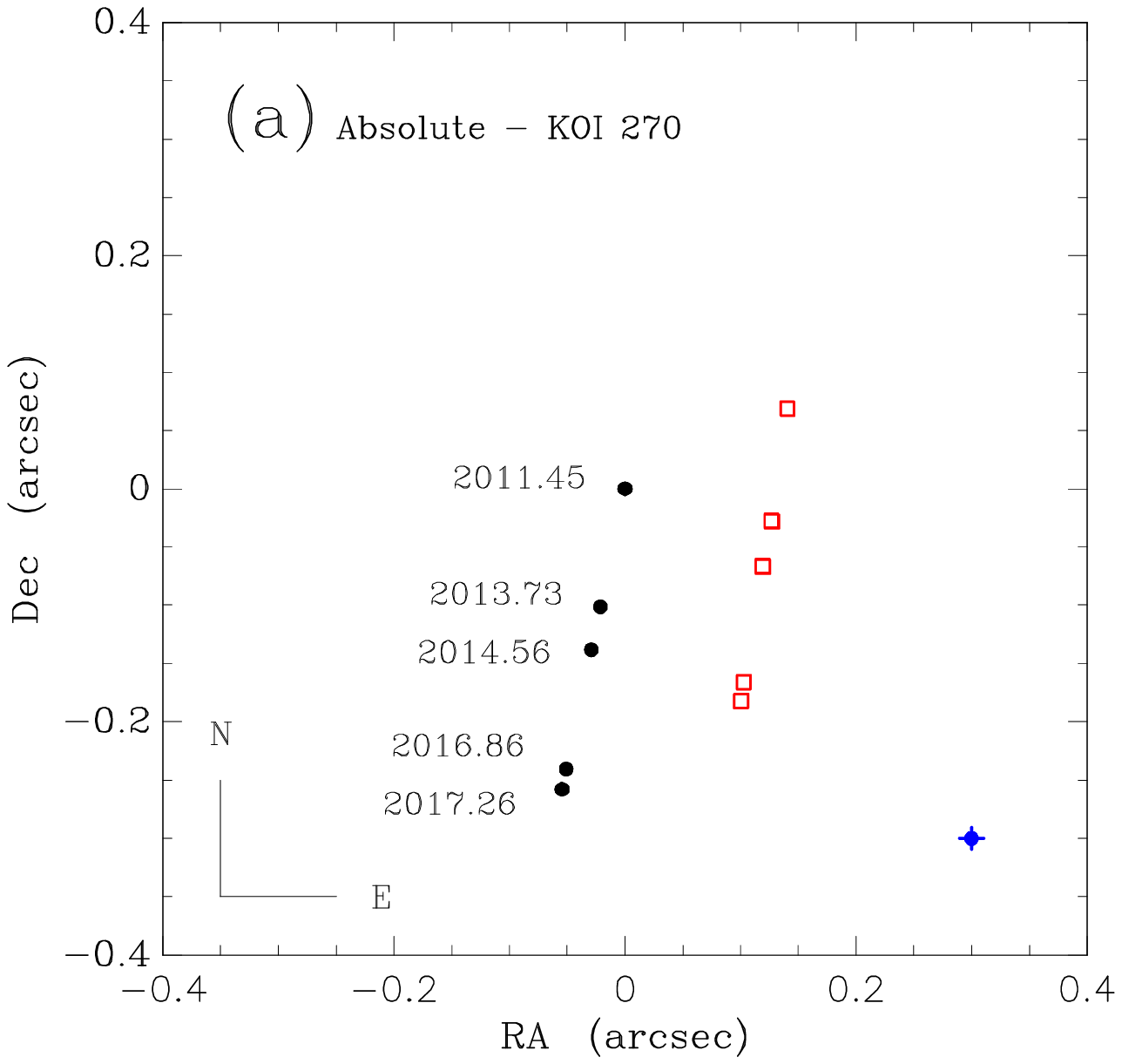}{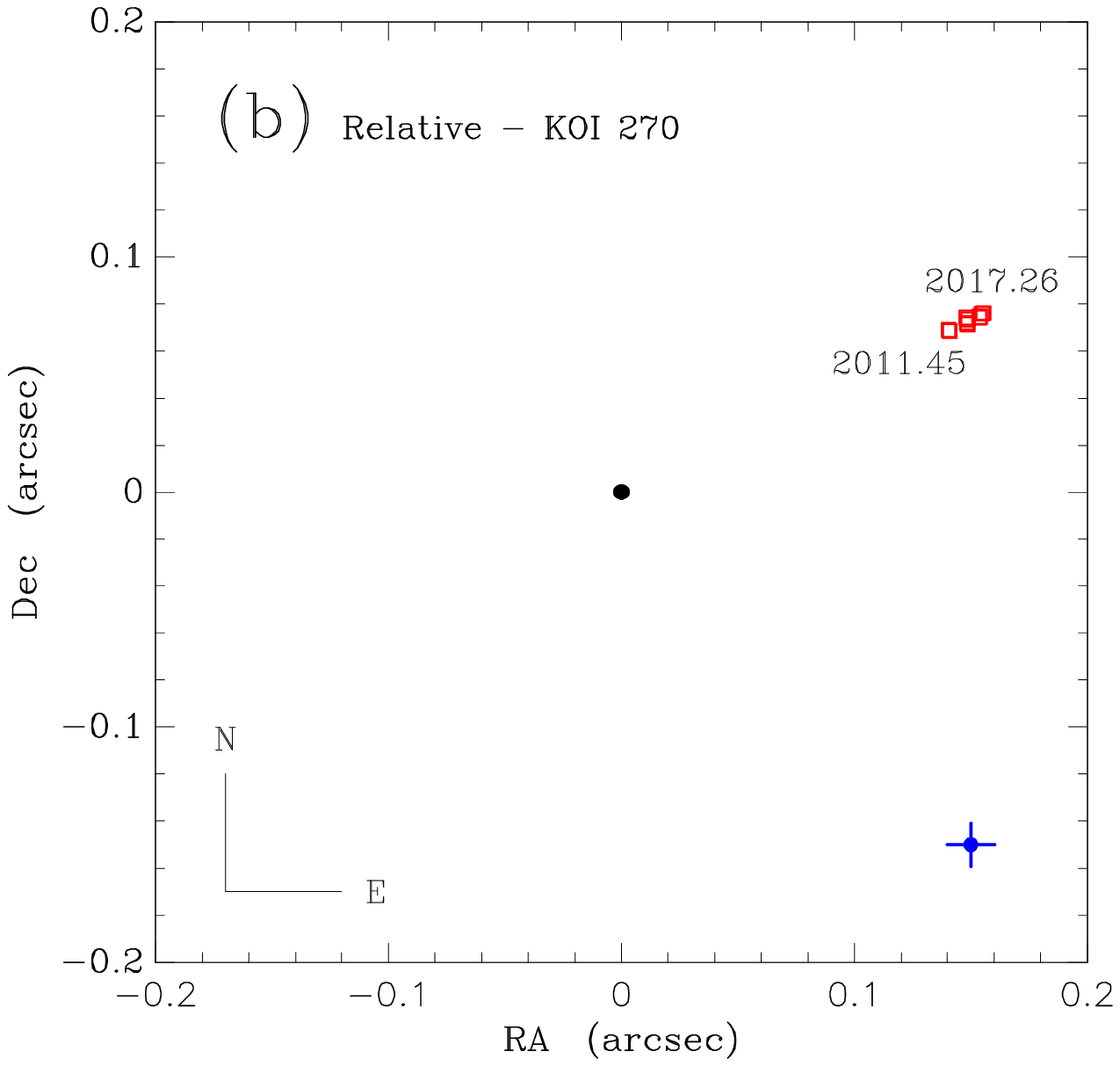}

\plottwo{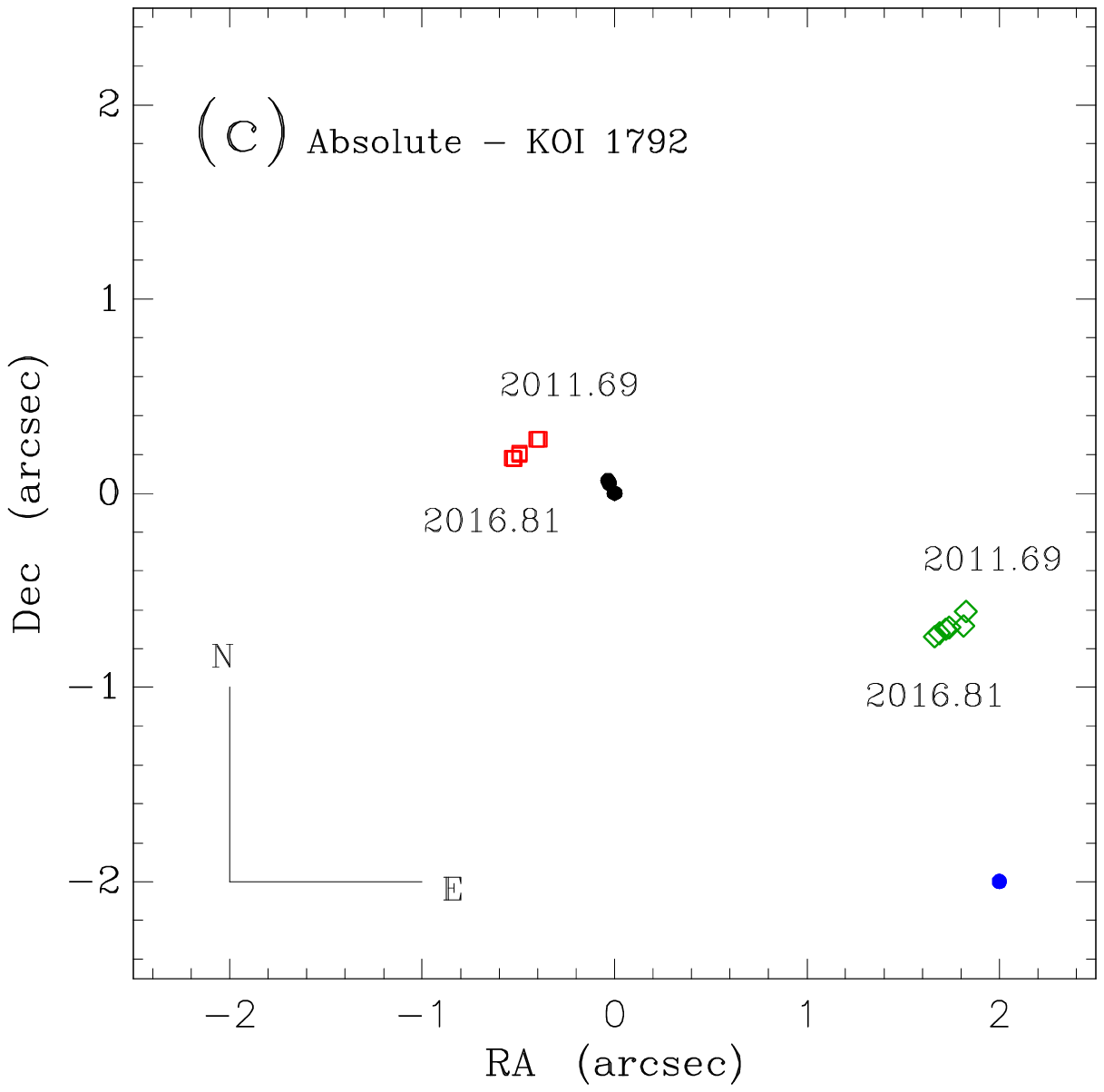}{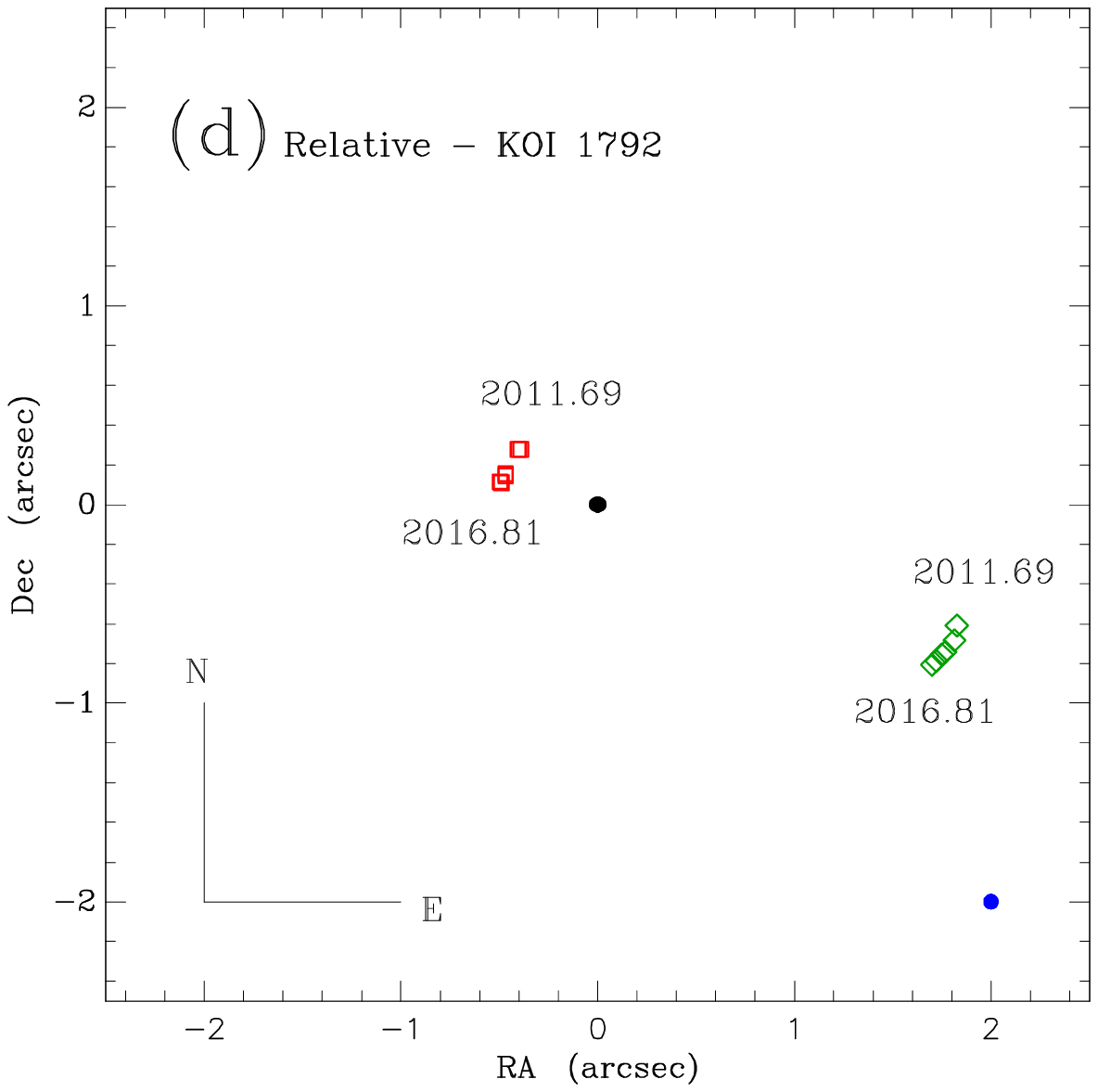}
\caption{
Absolute and relative motions of KOI 270 (Kepler-449, top) and KOI 1792 (Kepler-953, bottom). 
In all cases, the 
solid circles represent the location of the primary star, red squares indicate the secondary, and
green diamonds represent the tertiary. The blue point in the lower right indicates the uncertainty
of the total distance traveled over the period of time shown; it is the uncertainty in the proper motion
multiplied by the time baseline. (a) The absolute motions of the
two stars in the KOI 270 system ; this is judged to be a common proper motion pair. (b) The relative
motion of the secondary star in KOI 270 relative to the primary. The motion seen here is consistent
with an edge-on orbit at this point, though the separation is currently increasing and the likelihood of 
a long orbit is high if the system is bound. (c) 
The absolute motions of the triple star KOI 1792. (d) The relative motions of 
the secondary and tertiary stars with respect to the primary for KOI 1792. Here we see that the
fainter pair is moving together, in a direction nearly orthogonal to the system proper motion.}
\end{figure}

\vspace{0.2cm}
In Figure 8, we show a further visualization of the motions of 
KOI 270 (Kepler-449) and KOI 1792 (Kepler-953). In panel (a) we show the motion on the 
sky for KOI 270, as derived from the system proper motion and the relative positions over time. In
panel (b) the motion of the secondary star relative to the primary is plotted. Panels (c) and (d) show
the same two plots for KOI 1792. In the case of KOI 270, the system has a relatively large proper
motion, which clearly dominates over the relative motion. Nonetheless, when the relative motion is plotted, the 
motion of the secondary is clearly revealed to be in a direction radially away from the primary. 
In the case of the triple star
KOI 1792, we see that the secondary and tertiary stars are on opposite sides of the primary, yet
they move together in a direction nearly orthogonal to the (small) proper motion of the system
as a whole. Thus, the second and third components form a co-moving pair that is not related
to the central star.

\section{Discussion}

With a future analysis carried out over a longer time baseline, 
it may be possible to see orbital motion develop in some systems studied here. Regular observations
of these systems would have to be done over at least the next decade and a rigorous astrometric calibration 
regimen would need to be maintained, but for the fastest moving cases, such a program may
yield evidence of deviations from linear relative motion.
The motions detected so far on our sample of stars do not show convincing evidence of orbital 
acceleration, but this is not surprising given projected separations, which are generally in the range of
100s of AU. Nonetheless, we did attempt to find and examine the most likely cases where orbital
motion may reveal itself in the coming years. To do this, we consider the region of Figure 7
in the lower left part of the plot, where the projected separations are the smallest and the masses
of the components will be comparable. Cross-referencing those systems with their distances and 
selecting those with the smallest distances (so that the angular separation represents a smaller
physical separation), 
we find that two systems that are most worthy of future study in this regard are KOI 270 (Kepler-449)
and KOI 1613 (Kepler-907), both of which have current projected physical separations between 
40 and 50 AU. In the former case, as already discussed, the motion observed so far is 
in a direction along the separation vector, as would be expected in the case of an edge-on system,
implying that this stellar companion's orbit is possibly co-planar with the orbit of the planet in 
that system. In the latter, the uncertainties in the relative proper motions preclude a definitive
statement on the direction of the secondary's motion relative to the primary at this stage. However, 
the values measured here suggest that it is not obviously consistent with an edge-on orbit, and if
that were to be the case, it would represent an orbit that cannot be
co-planar with the planet in that system. At least several more years of data will 
be needed to determine if these hypotheses are correct. Two other systems with confirmed
exoplanets, KOI 307 (Kepler-520) and KOI 2059 (Kepler-1076), have current projected separations
in the range of 50-100 AU, and will also be worth continued monitoring over the next few years.

\begin{figure}[!t]
\figurenum{9}
%\plottwo{f9.eps}
%\figcaption{
\hspace{4.5cm}
\includegraphics[scale=0.50]{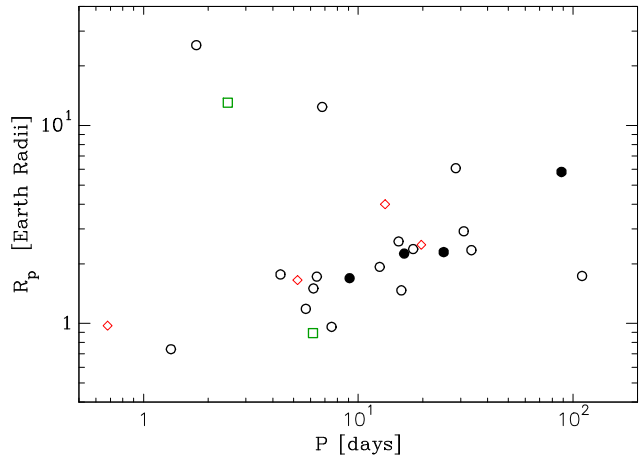}
\caption{
A period-radius plot for the confirmed planets in systems included in this study. Open circles are used
for the systems in Table 3 judged to be CPM pairs, green squares indicate 
probable-CPM pairs, filled circles are drawn for LOS companions, and red diamonds indicate 
those systems for which no determination can be made at present.}
\end{figure}

In Figure 9, we show a period-radius relation for the confirmed exoplanets orbiting the stars in our 
current sample.
The data are again taken from the Kepler CFOP website, where the radius values were then
corrected for the dilution of the secondary star and the planet is assumed to orbit the primary.
As \citet{ful17} and other subsequent papers
have shown, this plot contains a wealth of information relevant for planet formation and evolution. 
We have small sample statistics at present and the speckle observations do not represent a complete
sample where observational biases can be fully accounted for; therefore, it would be unwise to over-interpret the plot as it stands, but a few initial comments can be made. First, the basic features of this plot, such as 
the major branch of data points following a trend toward smaller planetary radii as the period decreases
and a lack of intermediate-sized planets with periods of less than 10 days, appear similar to the much 
larger sample of exoplanets orbiting single stars. Second, we do not yet have examples of planets in binary 
systems with radii exceeding 25$R_{\oplus}$, whereas the current known sample of exoplanets contains a well-defined sample of giant planets with radii at least as large as 40-50 $R_{\oplus}$. In an examination of planetary periods and radii with stellar properties such as mass and metallicity, no obvious trends were identified in our sample.

As with most of the \textit{Kepler} discoveries, the planetary orbital periods are relatively short. Given the 
generally large separation of the binary pairs ($>$ 100 AU), these short-period planets are not likely to be 
greatly influenced by the presence of the stellar companion in the system. However, the planetary systems 
Kepler-449 and Kepler-907 are of potential interest in this regard due to the possibility of detecting
orbital motion in the coming years. These planetary systems would be the most obvious candidates 
in this study with which to search for any 
dynamical effects the close stellar pairs may have on the currently known planets. Orbital resonances, inclination effects, or eccentric orbits may provide clues to gravitational influence by the bound companion, once enough high-quality astrometric data is obtained. Theoretical limits on additional outer planetary orbits might be explored as well.

The planets contained in large-separation binaries are all likely to orbit the primary star 
\citep{2018AJ....155..244B}, and the majority of the confirmed planets in our systems are of size 1-3 
R$_{\Earth}$. Kepler-1, -13, and -14 contain large, ``Hot Jupiters" in close orbits, with migration possibly influenced by the bound companion. One of our CPM binary  pairs, Kepler-449, has a motion consistent with an edge-on orbit, and would therefore be co-aligned with that of its planetary system. Continued observations of such systems, as well as others discovered by the {\it K2} and {\it TESS} missions, is an important step in fully characterizing the motions and eventually determining binary orbits. That
would give a unique viewpoint on understanding binary star and exoplanet system formation and evolution.

\section{Conclusion}
We have presented relative astrometry and photometry measurements obtained from speckle imaging 
for 57 KOIs over the time frame of 2010 to 2018, and performed a detailed study of 34 objects
 from the sample for which we have sufficient data to measure the relative proper motions of the system. 
 Three of these systems are triple stars, and 18 of these systems have at least one confirmed exoplanet.
 Of those 18 exoplanet systems, we judge 12 to be either CPM or probable-CPM systems, 
 and 2 to be LOS pairs.
 Four are of uncertain disposition based on our current measurements. Orbital periods for the confirmed 
 planets in our sample range from 0.7 to 110 days, with a median period of 7.5 days. Our results are generally
 in agreement with the photometric analysis of \citet{2017AJ...153...117H} where both studies
 were able to make determinations; the two exceptions
 are cases where the astrometry suggests a common proper motion but the earlier photometric analysis
 was not able to put the two stars on a common isochrone.
 
A preliminary period-radius relation for the confirmed exoplanets that are known in our sample reveals
trends that comparable to the diagram for all exoplanets, including a lack of planets of radii comparable
to Neptune at periods less than 10 days. This sample of {\it Kepler} CPM pairs hosting exoplanetary 
systems serves as a starting point for observational and theoretical development and speaks to the need for 
additional long-term astrometric observations. While the typical \textit{Kepler} exoplanet host star is far away 
($\sim$800 pc), we suggest
that the same methodology presented here will provide a key tool that can be used on e.g.\ {\it TESS} stars in
future years. As those stars are generally are much closer to the Solar system, shorter-period binary exoplanet 
host stars would be much more common and orbital motion would presumably be much more rapidly detected 
in those cases, and more robust statistics of a wider variety of exoplanet hosts may be obtained.

\acknowledgements 
We are grateful to the excellent staff at the three observatories used in this project: WIYN, LDT, and Gemini-N. 
Our colleagues Gerard van Belle, William Sherry, David Ciardi, Johanna Teske, Lea Hirsch, 
Nic Scott, and Rachel Matson also 
participated in observing runs where some of these observations were taken, 
along with observations for other projects including their own. All of these individuals
made our observing work more 
productive and enjoyable with their high degree of collegiality and professionalism. 
We also thank William van Altena for his 
helpful comments on the astrometric methods presented here. N.M.C. is grateful for a Southern Connecticut
State University Undergraduate Research Grant to complete this work. E.P.H. thanks the National Science 
Foundation for NSF Grants AST-0908125 and AST-1517824, which allowed him to participate in the 
observing for this project. 
J.W.D. gratefully acknowledges funding from NASA through two PI Data Awards administered by the NASA 
Exoplanet Science Institute. 

This work was based in part on observations obtained at the Gemini Observatory, which is operated by the 
Association of Universities for Research in Astronomy, Inc., under a cooperative agreement with the NSF on 
behalf of the Gemini partnership: the National Science Foundation (United States), National Research Council 
(Canada), CONICYT (Chile), Ministerio de Ciencia, Tecnolog\'{i}a e Innovaci\'{o}n Productiva (Argentina), 
Minist\'{e}rio da Ci\^{e}ncia, Tecnologia e Inova\c{c}\~{a}o (Brazil), and Korea Astronomy and Space Science 
Institute (Republic of Korea). As visiting astronomers to Gemini-N, we are mindful that Maunakea is a sacred 
space to many native Hawai`ians, and we are grateful for the opportunity to have been present there. Likewise, 
Kitt Peak is home to the Tohono O'odham people, and we are privileged to have been visitors to 
that special place.

Some of the observations in the paper made use of the NN-EXPLORE Exoplanet and Stellar Speckle Imager 
(NESSI). NESSI was funded by the NASA Exoplanet Exploration Program and the NASA Ames Research 
Center. NESSI was built at the Ames Research Center by Steve B. Howell, Nic Scott, Elliott P. Horch, and 
Emmett Quigley.



%\figsetgrpstart
%\figsetgrpnum{1.7}
%\figsetgrptitle{V2672 Oph}
%\figsetplot{V2672_Oph-eps-converted-to.pdf}
%\figsetgrpnote{The Swift/XRT X-ray light curve for the first year after
%outburst.}
%\figsetgrpend

%\figsetgrpstart
%\figsetgrpnum{1.8}
%\figsetgrptitle{V407 Cyg}
%\figsetplot{V407_Cyg-eps-converted-to.pdf}
%\figsetgrpnote{The Swift/XRT X-ray light curve for the first year after
%outburst.}
%\figsetgrpend

%\figsetend

\end{document}